\documentclass{elsart5p}
\usepackage{graphicx}
\usepackage{amssymb}
\usepackage{amsmath}
\usepackage{multirow}

\begin{document}

\title{Void Nucleation, Growth, and Coalescence in Irradiated Metals}
\author{Michael P. Surh, Jess B. Sturgeon$^\dagger$, and Wilhelm G. Wolfer}
\address{Lawrence Livermore National Laboratory, 7000 East Ave., Livermore, California, 94551, USA; $^\dagger$ Perceptive Software, Shawnee, KS 66226
}

\maketitle

\section{Abstract}
A novel computational treatment of dense, stiff, coupled reaction rate equations is introduced to study the nucleation, growth, and possible coalescence of cavities during neutron irradiation of metals. 
Radiation damage is modeled by the creation of Frenkel pair defects and helium impurity atoms.
A multi-dimensional cluster size distribution function allows independent evolution of 
the vacancy and helium content of cavities, distinguishing voids and bubbles. 
A model with sessile cavities and no cluster-cluster coalescence can result in a
bimodal final cavity size distribution  with coexistence of small, high-pressure bubbles and large, low-pressure voids.
A model that includes unhindered cavity diffusion and coalescence ultimately removes the small helium bubbles from the system, leaving only large voids.
The terminal void density is also reduced and  the incubation period and terminal swelling rate can be greatly altered by cavity coalescence.
Temperature-dependent trapping of voids/bubbles by precipitates and alterations in void surface diffusion from adsorbed impurities and internal gas pressure may give rise to intermediate swelling behavior through their effects on cavity mobility and coalescence.

\section{Introduction}

Irradiation of metals has long been known to culminate in macroscopic property changes including void swelling \cite{CAWTHORNE:1967}.
Characteristic stable voids and steady volumetric swelling develop for a range of temperatures and fluxes, independent of whether radiation bombardment damage occurs as disseminated Frenkel pairs or as small defect clusters.  
This can occur whether or not helium is generated along with atomic displacements.
In either case, small, unstable voids, loops, and other defect clusters will develop almost immediately within the irradiated material.  
Their subsequent evolution determines the fluence required to create stable voids and achieve steady swelling; this so-called incubation dose includes most of the dependence on radiation environment \cite{GarnerWolfer:1984,Okita:2000,Okita:2002}.
The processes that govern microstructure evolution include thermally-activated motion of small defect clusters, mutual impingement, and annihilation or coalescence reactions
along with micro-chemical changes from nuclear transmutation and displacements or diffusion of pre-existing impurities.  
Radiation simulations should ideally encompass all of these processes.  
Typically, existing models have included only particular types of defects and reactions or have made other numerical approximations in order to obtain a solution.

At the least, simulations of early irradiation must account for void nucleation and growth processes, since annihilation,  aggregation, and cluster ripening take place concurrently.  
Transient and steady-state swelling behavior due to these processes have been studied recently \cite{Surh:2004,Surh:2004b,Surh:2005,Surh:ERR}.
However, only void reactions with vacancy or interstitial monomers are included in these studies. 
This minimal model of void nucleation gives reasonable swelling behavior as a function of temperature and flux \cite{Surh:2005,Surh:ERR}, viz. an observed steady swelling rate around 1$\%$/dpa in austenitic stainless steels  and an important flux-effect on the measured incubation times
 \cite{Garner:1998,Okita:2001b}.
While the results are encouraging, these calculations neglect many of the processes believed to shape the microstructure.  
For example, the generation and aggregation of helium impurities is not explicitly modeled.  
Size-dependent void diffusion \cite{GREENWOOD:1963,GRUBER:1967} is neglected, and thus direct  void-void coalescence is not included.
Dislocation loop formation, migration, and coalescence is not explicitly simulated, either.
The model can be considered to combine the production and biased diffusion of small vacancy and interstitial clusters into {\it effective} generation and reaction rates for monomer species alone, but it is unclear a priori how a coarse-grained treatment of these processes affects microstructure evolution.

It is now clear that the model must presuppose a ready supply of gas impurity atoms (e.g., oxygen and helium \cite{COGHLAN:1983}) to promote the formation of small voids from the radiation-induced, supersaturated vacancy population.  
In practice, reasonable corrections to void energies  may reproduce the approximate void number density observed in irradiated steel \cite{Surh:ERR}.  
Ultimately, however, crude models for the apportionment of impurities among the defect clusters should be supplanted by a detailed accounting of multicomponent aggregation and coalescence reactions and their influence on the non-equilibrium cluster size distribution.
Such problems are widely addressed in the literature, including gelation, polymerization, and formation of aerosols and precipitates in solid or fluid media \cite{MARCUS:1968,GILLESPIE:1972,LUSHNIKOV:1978,VODE,SMITH:1998,BABOVSKY:1999,GILLESPIE:2000,GILLESPIE:2001,EIBECK:2001,HASELTINE:2002,FRIESEN:2003,LAURENZI:2003,MUKHERJEE:2003,ALEXOPOULOS:2004,FILBET:2004,SALIS:2005,KRAFT:2005}.
The numerical methods developed for such problems may also be fruitfully applied to radiation swelling.
Here, a hybrid numerical approach that can encompass the full range of possible cluster compositions and cluster reactions in mean field is introduced, a Livermore Microstructure Evolution program, {\it LiME}.
As a first application, the method is applied to the nucleation and growth of voids with a two-component  distribution of cluster compositions, examining the evolution of helium-vacancy clusters \cite{COGHLAN:1983}, while continuing to treat oxygen adsorption by simply reducing the cavity surface energy by a constant (temperature-independent) factor.
The method predicts realistic swelling behavior for ferritic steel in reactor environments.

As before, the void distribution function is partitioned into overlapping regions \cite{Surh:2004}, treating small clusters with the Master Equation (ME domain) and large ones with Monte Carlo methods (MC domain).  
This allows self-consistent evolution of the full void population with no truncation of the size domain, 
no assumptions as to the critical void size or the nature of the nucleation process,  and no approximations for the overall nucleation rate or duration of the nucleation stage.  
Monomer concentrations are included in the ME region, where they may either be treated separately by a quasi-stationary approximation or evolved along with the small clusters through coupled nonlinear reaction rate equations.
The formation and evolution of dislocation loops is not explicitly modeled;  
network dislocations and loops are already described by a single, time-dependent density parameter rather than a detailed size distribution function \cite{WOLFER:1985}.
However, the methods used for void evolution would be easily generalizable to other defect species and reactions, provided that suitable mean field rate coefficients are specified for their reaction rate equations.
In particular, future calculations will consider the formation, unfaulting, and migration of dislocation loops;  loop coalescence and annihilation; and incorporation of loops in the dislocation network.

The remainder of this paper first describes the coupled, stiff, non-linear evolution equations for void nucleation, growth, and coalescence.
It  presents the microscopic rate theory model, gives an overview of the computational scheme,  details
the various numerical methods employed in the calculations, and
makes a preliminary application to void nucleation in irradiated stainless steel.
The simulations include vacancy, interstitial, and helium generation, aggregation and, annihilation, with or without cluster coalescence.
The results are sensitive to the effects of absorbed impurity atoms on cavity surface energy.
They also expose a substantial influence of small, unstable, {\it mobile} clusters on the formation of critical-sized voids via direct cluster-cluster coalescence.
Realistic incubation and swelling behavior cannot be obtained over wide ranges of temperature and flux without including cluster mobility and coalescence.

\section {Rate Theory Model}

\label{RTM}

Allowable microstructure reactions (either aggregation or annihilation) are assumed to occur whenever two defects, ${\mathbf m}$ and ${\mathbf n}$, come into contact.  
Within the mean field continuum approximation, the collision rate is proportional to their relative diffusivity, $D_{{\mathbf m},{\mathbf n}}$, and effective collision cross-section, $A_{{\mathbf m},{\mathbf n}}$.
As before \cite{Surh:2004,Surh:ERR},  a bias factor $Z_{{\mathbf m},{\mathbf n}}$ includes the effect of long-range interactions \cite{SNIEGOWSKI:1983,SurhWolfer:TBP}
on the binary reaction rates, $K({{\mathbf m},{\mathbf n}}) \rho_{\mathbf {\mathbf m}} \rho_{\mathbf n}$, 
where $\rho$ are densities of reactant species ${\mathbf m}\ne {\mathbf n}$ and the rate coefficients are:
 \begin{equation}
K({{\mathbf m},{\mathbf n}}) = Z_{{\mathbf m},{\mathbf n}} A_{{\mathbf m},{\mathbf n}} D_{{\mathbf m},{\mathbf n}} 
\label{KEQN}
\end{equation}
Note that an additional factor of 1/2  may be required when ${\mathbf m}={\mathbf n}$, to prevent double-counting of unique pairs of identical reactant particles. 
This factor is not explicitly shown in the definition of $K$.  

Microstructure defect species are limited here to self-interstitials and -vacancies, substitutional and interstitial helium, voids/bubbles, and network dislocations.
Vacancy and helium monomers as well as clusters are characterized by their composition, ${\mathbf n}=(n_{vac},n_{hel})$, in a two dimensional space.
Self-vacancies and interstitials are also specially identified by $(1,0)=v$ and $(-1,0)=i$, respectively; 
substitutional and interstitial helium by $(1,1)=vh$ and $(0,1)=h$;  and network dislocations by $d$.
Monomer densities evolve according to:
\begin{align}
{{d\rho_i}\over{dt}} = g_i 
& -\sum_{\mathbf m \not\in\{h,i\} } K({\mathbf m},i)   \rho_{\mathbf m} \rho_i -  K(d,i) \rho_{d}\rho_i \cr
{{d\rho_h}\over{dt}} = g_h   
& -\sum_{{\mathbf m \not\in\{ h,i\} }} K({\mathbf m},h) \rho_{{\mathbf m}}\rho_h + K(vh,i) \rho_{vh} \rho_i\cr
{{d\rho_v}\over{dt}} = g _v  
& - \sum_{{\mathbf m }} K({{\mathbf m},v}) \rho_{\mathbf m} \rho_v 
+ \sum_{\mathbf m} \bigl(K({{\mathbf m}-v,v}) c_v^{[{\mathbf m}]}\bigr) \rho_{\mathbf m} \cr
&  +K({v_2,i}) \rho_{v_2} \rho_i - K({d,v}) \rho_d\rho_v  + (K({d,v}) c_v^{[eq]}) \rho_d  \cr
{{d\rho_{vh}}\over{dt}} = g_{vh}  
&
-\sum_{\mathbf m}
K({{\mathbf m},vh})  \rho_{\mathbf m} \rho_{vh}
+K({v,h})\rho_{v} \rho_{h}
\cr
&+ (K({vh,v}) c_v^{[v_2h]})\rho_{v_2h}
+ K({v_2h,i}) \rho_{v_2h} \rho_i  
\label{MEQN}
\end{align}
The vacancy-vacancy aggregation term, $(Z_{v,v} A_{v,v} D_{v,v}) \rho_v \rho_v$,  within the first summation for ${d\rho_v}/{dt}$ in Eq.~\ref{MEQN} includes that  two vacancies are consumed by the reaction, that there is a factor of 1/2 to prevent double-counting of unique pairs of vacancies from the population $\rho_v$, and that the relative diffusivity is twice $D_v$.  
The net rate is identical to that used in a previous study \cite{Surh:2004}.
Similar considerations also apply to pairs of substitutional helium and to thermal dissociation of vacancy dimers.

Cluster (${\mathbf n}\not\in \{ v,vh,h,i\}$) densities evolve as:
\begin{align}
{{d\rho_{\mathbf n}}\over{dt}} =   g_{\mathbf n} 
&
  + \biggl \{ \sum_{{\mathbf m}\in(v,vh,h)}
   K({{\mathbf m},{\mathbf n}-{\mathbf m}}) 
    \rho_{\mathbf m} \rho_{{\mathbf n}-{\mathbf m}} 
  \enskip \bigl (1-{\delta_{\mathbf m,\mathbf n-\mathbf m}\over2}\bigr)
    \bigl (1-U({\mathbf m}-{\mathbf n}) \bigr)
 \cr
 &
-\sum_{{\mathbf m}\in(v,vh,h)}
   K({{\mathbf m},{\mathbf n}}) 
     \rho_{\mathbf m} \rho_{{\mathbf n}}
\cr
&
   +K({{\mathbf i},{\mathbf n}+v})
    \rho_{\mathbf i} \rho_{{\mathbf n}+v} 
   -K({{\mathbf i},{\mathbf n}}) 
     \rho_{\mathbf i} \rho_{{\mathbf n}}
     \enskip  U(\mathbf n-v)
\cr
&
  +(K({{\mathbf n},v}) c_v^{[{{\mathbf n}+v}]})\rho_{{\mathbf n}+v} 
   -(K({{\mathbf n}-v,v}) c_v^{[{\mathbf n}]})\rho_{{\mathbf n}}
     \enskip U(\mathbf n-v)
\biggr\}
\cr
&+\biggl \{
- \sum_{{\mathbf m}\not\in(i,v,h,vh)} 
K({{\mathbf m},{\mathbf n}}) 
  \rho_{\mathbf m}\rho_{\mathbf n} 
\cr
&
+ {1\over2}{\sum}^\prime 
K({{\mathbf m},{\mathbf n-\mathbf m}})   \rho_{\mathbf m} \rho_{{\mathbf n}-{\mathbf m}}  
\bigl (1-\enskip U({\mathbf m}-{\mathbf n})\bigr )
\biggr\}
\label{RATEEQN}
\end{align}
in terms of any direct generation of clusters in the radiation damage cascade, $g_{\mathbf n}$; reactions of existing clusters with monomers (in brackets) that consume or create
${\mathbf n}$-mers including thermal emission of vacancies, and cluster-cluster reactions (in the second set of brackets) that consume or create ${\mathbf n}$-mers. 
Factors of 1/2 in the first and last summations prevent double counting of indistinguishable 
reactant pairs,
and $\delta_{\mathbf m,\mathbf n} = \delta_{m_{v},n_{v}}\delta_{m_{h},n_{h}}$ where the right hand side consists of the usual Kronecker deltas,
$\delta_{i,j}=\begin{cases}1& i= j\\0&i\ne j \end{cases}$.
The primed summation is restricted to all pairs of reactants with ${\mathbf m}, {\mathbf {n-m}} \not\in\{v,vh,h,i\}$.
Defects $\mathbf n - \mathbf m$ ($\mathbf n - v$, etc.)  are restricted to the  domain of allowed compositions by a step function: $U({\mathbf n}) = U(n_{v})U(n_{h})$, where $U(n)=\begin{cases}1& n\geq0\\0&n<0\end{cases}$. 
Finally, clusters never undergo fission in this model, only the thermal emission of single vacancies.  

Radiation damage deposition is approximated by the creation of disseminated monomers, 
so $g_{{\mathbf n}}\equiv0$ for ${{\mathbf n}}\not\in\{v,vh,h,i\}$. 
In this case, $g_i = \phi \xi$,  in terms of the atomic displacement rate, $\phi$, and the damage production efficiency, $\xi$. 
The total helium production is $g_h+g_{vh}$, with
the ratio of interstitial to substitutional depending on the model.
(Here, it is assumed that the helium is all deposited as substitutional defects.)
Conservation of host atoms (including transmutation products) requires $g_v + g_{vh}  \equiv g_i$.  
Helium impurities are added with a temperature-independent, gradual activation of $\alpha$-emitters. 
This is modeled for a Fe-Ni-Cr steel undergoing neutron bombardment according a two-step activation process, in analogy to the $^{58}$Ni(n,$\gamma$)$^{59}$Ni(n,$\alpha$) reaction.
Model transmutation rates are treated as free parameters and are fit to the experimental helium content in HFIR-irradiated nickel \cite{GARNER:2003,SCHALDACH:2003}.
The parameters are $\gamma$, $\alpha$, and $\delta$ for the rates of (respectively) $^{58}$Ni(n,$\gamma$), $^{59}$Ni(n,$\alpha$), and the sum of all transmutations that consume $^{59}$Ni.
In terms of  the cumulative radiation dose in dpa, $x=\int \phi(t) dt$ (for radiation flux, $\phi$):
\begin{equation}
{d\over{dx}}\left(
\begin{array}{c}
 \rho _{58} \\
 \rho _{59}
\end{array}\right)
\text{=}
\left(\begin{array}{cc}
 -\gamma  & 0 \\
  +\gamma & -\delta
\end{array}
\right)\left(
\begin{array}{c}
 \rho _{58} \\
 \rho _{59}
\end{array}
\right)
\label{EQ3}
\end{equation}
The $^{59}$Ni content, $\rho_{59}$, is obtained from Eq.~\ref{EQ3} by transforming to the eigenbasis,
where $\rho_A(x) = \rho_{58}(x)$ and
$\rho_B(x) =  \rho_{59}(x) + {\gamma\over{\gamma-\delta}} \rho_{58}(x)$ are solved, 
and then transforming back. 
The helium generation rate is given by:
\begin{equation}
{{d  \rho_{He}}\over{d x}} = \alpha \rho_{59}(x) = \alpha  {\gamma\over{\gamma-\delta}} \rho_{58}(0) \bigl [ e^{-\delta x} - e^{-\gamma x} \bigr ] 
\end{equation}
assuming that only $^{59}$Ni(n,$\alpha$) produces $\alpha$-particles.
The fit parameters are $\gamma=0.0255$, $\alpha=0.0711$, and $\delta=0.297$ dpa$^{-1}$.
Pristine type 316 stainless steel  is approximately 14$\%$ nickel, with 68.08\% of that $^{58}$Ni and with no naturally-occurring  $^{59}$Ni.
Other relevant materials parameters  for type-316 stainless steel are listed in Table~\ref{TableOne}.

Non-interacting diffusion (independent random walks) implies $D_{\mathbf m,\mathbf n} = D_{\mathbf m}+D_{\mathbf n}$.  Defect collision cross-sections are simply given by
\begin{align}
A_{{\mathbf m},{\mathbf n}} &= 4\pi(r_{\mathbf m} + r_{\mathbf n}) &{\rm for }\enskip {\mathbf m}\not\in\{v,vh,h,i\} \enskip {\rm and } \enskip {\mathbf n}\not\in\{v,vh,h,i\}\cr
A_{{\mathbf m},{\mathbf n}} &= r_{\mathbf m} + b &{\rm for} \enskip {\mathbf n}\in\{v,vh,h,i\}
\label{XSection}
\end{align}
in terms of radii for (spherical) defects, $r_{\mathbf n}=\sqrt{ {(3n_{v}\Omega)^2}\over{4\pi} }$
(except for interstitial monomers, where $r_i=r_h=r_v$).
For consistency with earlier work, cross-sections involving monomers are defined using the Burgers vector magnitude in place of a monomer radius. 

Bias factors between voids and the four defect monomers are calculated  from a mean field solution of the diffusion including stress-mediated interactions  \cite{SurhWolfer:TBP}.  
The infinite series describing the image interaction \cite{MOONPAO:1967} is fit by a simple analytic form,  while the modulus interaction \cite{WOLFERASHKIN:1973} is treated analytically. 
The numerical results are 
tabulated for small voids and computed as needed for larger ones.
Long range void-void interactions are presently neglected, so $Z_{{\mathbf m},{\mathbf n}}=1$ for ${\mathbf m},{\mathbf n}\not\in\{v,vh,h,i\}$.
In principle, the effect of any  long-range interactions or net drift velocities (e.g., from external stress or temperature gradients \cite{COTTRELL:2002}) can be incorporated in the void-void reaction rates, so the mean field reaction kernel, $K$, has general applicability.

Thermal emission from vacancy clusters is evaluated by a detailed balance argument.
Equating vacancy emission and absorption for the ${\mathbf n}$-mer  identifies the chemical potential, $\mu_v^{[{\mathbf n}]} = F^{[{\mathbf n}]} - F^{[{\mathbf n-v}]}$, in terms of the $\mathbf n$-mer and $({\mathbf n}-v)$-mer (i.e., void minus one vacancy) free energies.
Rewriting in terms of void internal energies, $E$, and the inert gas pressure, $P$:
\begin{equation}
c_v^{[\mathbf n]}= c_v^{[eq]} e^{(E^{[{\mathbf n}]}-E^{[{\mathbf n}-v]}-P\Omega)/kT}
\label{CVEQN}
\end{equation}
Gas pressure is described with a non-ideal equation of state for helium versus density and temperature \cite{Wolfer:1988}.  
No volume relaxation is included (i.e., the void volume is  $n_{v}\Omega$).  
In the absence of surface-adsorbed impurity atoms, $E^{[{\mathbf n}]}$  is parametrized in terms of an effective surface energy, $\gamma^{[{\mathbf n}]} $, and the surface area of a spherical cavity of volume $n_v \Omega$
\begin{equation}
E^{[{\mathbf n}]} 
= \gamma^{[{\mathbf n}]} 4\pi r_{\mathbf n}^2 =  \Lambda\gamma^0(T) \biggl( 1 - {0.8\over{n_v+2}}\biggr)4\pi r_{\mathbf n}^2
\label{SIZDE}
\end{equation}
In the continuum limit,  $\gamma^{[{\mathbf n}]} $
approaches that of a flat, clean surface, $\gamma_0(T)$, while it approaches
the results of atomic calculations in the limit of small voids \cite{ADAMS:1989}.
This surface energy is then further reduced by an constant scale factor, $\Lambda$, to reflect the presence of adsorbed oxygen impurities \cite{English:1987} (see Table~\ref{TableOne}).
Finally, the emission rate is obtained from $c_v^{[{\mathbf n}]}$ and the vacancy-cluster reaction parameters for the 
$({\mathbf n}-v)$-mer.
For straight, jogged dislocation segments,  $c_v^{[d]} =c_v^{[eq]}$, the thermal equilibrium concentration.
Emission rate coefficients in Eq.~\ref{MEQN} are represented as unary reactions, by including the defect-dependent $c_v^{[{\mathbf n}]}$ within the rate coefficient.

At some maximum density, an over-pressurized bubble would begin to emit self-interstitials via loop punching \cite{Wolfer:1988}.
Such a possibility is not considered here;  
instead,  an artificial constraint is imposed on the helium density in a reactant cluster, $n_{h}\leq 2 n_{v}$.
Any reactions that would yield a higher density are disallowed.  
Thermal dissociation of substitutional helium into a self-vacancy plus interstitial helium is similarly assumed to be energetically impossible at temperatures of interest.
Note that self-interstitial and interstitial helium aggregation is excluded
since interstitial loops are effectively part of the dislocation density model.
Mixed interstitial clusters can develop in principle \cite{WILSON:1983}. 
 
Void diffusivity $D_{\mathbf n}=D_v/{n_{v}}^{4/3}$  for ${\mathbf n}=(n_{v},n_{h})$.  
This gives both the correct monomer value and size-dependence for large cluster diffusion 
\cite{GREENWOOD:1963,GRUBER:1967}, although the activation energy for void 
migration should more properly be that for surface diffusion.
This diffusivity takes no account of the effect of reversible pinning \cite{NELSON:1966}, or internal gas pressure on the migration \cite{MIKHLIN:1979}, or radiation-enhanced diffusion \cite{ALEXANDER:1992}, or, e.g., that vacancy dimer diffusion may be $D_{v_2}\simeq D_v$.  
Trapping at dislocations and grain boundaries are not considered.
Such features would be straightforward to incorporate in the future.

The dislocation model reproduces measured dislocation densities versus dose and temperature \cite{WOLFER:1985}.
It includes separate source and annealing terms in terms of the biased flow of radiation-induced 
vacancies and interstitials.
There is one adjustable parameter, $l$, representing a characteristic dislocation pinning length \cite{WOLFER:1985}.
This is taken to be independent of the density of voids/bubbles in the matrix,
because the pinning length in stainless steels is more determined by carbide nano-precipitates than by the density of voids/bubbles.

\section{Numerical Method}
\subsection {Overview}

Once the temperature- and radiation-environment are specified and initial conditions for the microstructure are fixed,
the Master Equations~\ref{MEQN}~and~\ref{RATEEQN} completely determine the void/bubble size distribution function \hbox{$P(t)=\{ \rho_{\bf n}(t)\}$}. 
Such stiff, non-linear coupled rate equations can be integrated numerically \cite{VODE}, although this becomes intractable for a large domain of cluster sizes.
The number of distinct species may be reduced by grouping similar clusters together \cite{GOLUBOV:2001}, but the direct approach still becomes intractable for multi-dimensional distributions.    
Monte Carlo schemes for discrete coalescence events \cite{GILLESPIE:1972,GILLESPIE:2000} can naturally encompass large voids of arbitrary composition; however, they are inefficient for simulating nucleation from sub-critical clusters.
Here, the advantages of both methods are combined by partitioning  the 
cluster composition domain  into two overlapping regions.
Separate sub-distributions are defined for each,
labeled ME and MC for treatment by Master Equation and Monte Carlo, with $P = P^{ME} + P^{MC}$.
Each sub-distribution is composed of discrete ensembles of identical clusters, represented by $(\mathbf n,\rho)$ for the paired multi-dimensional cluster composition,  $\mathbf n$, and the ensemble density, $\rho$.  
The distribution $P^{ME}=\{ ({\mathbf n},\rho) \}^{ME}$  includes interstitials, 
$i$, and vacancy-helium clusters, $\mathbf n$,  with $0 \leq  n_{v} \leq N_{v}^{ME}$ and $0 \leq  n_{h} \leq N_{h}^{ME}$.  
There is exactly one element for each ME species, for a total of $N^{ME}$.
Only the densities of the ME elements evolve over time.
A sparse, random set, $\{( {\mathbf n},\rho)\}^{MC}$,  approximates $P^{MC}$
for  all $0 \leq n_{v}, n_{h} <\infty$.
The total number of elements, $N^{MC}$, is variable, and there may be none,  one, or many MC elements for a given ${\mathbf n}$ (each with potentially different values of $\rho$).
Both the densities and the compositions of the MC elements evolve with time.
Such split distribution functions have been used before in a Fokker-Planck treatment 
of void growth \cite{Surh:2004}, and in non-equilibrium chemistry \cite{HASELTINE:2002,SALIS:2005}, and plasma physics applications 
\cite{SOLOVYEV:1999}.  
In essence, the elements of $P^{MC}$ also constitute so-called "macroparticles",
already in wide use for non-equilibrium plasma physics problems \cite{MACRO}.

ME-ME reactions (those processes with reactants and product among the elements of $P^{ME}$)  are evaluated in a continuum approximation, using the Master Equation \cite{VODE}.
Discrete MC-MC reactions are performed stochastically using a Markov Monte Carlo procedure \cite{GILLESPIE:1972}.
ME-MC cross-reactions are included using either the Markov Monte Carlo method or Poisson-distributed random walks \cite{GILLESPIE:2000,GILLESPIE:2001} for $P^{MC}$, and using average sink or source terms in the rate equations for $P^{ME}$.  
There are also procedures to transfer clusters between the two sub-distributions and to regulate the number of elements and their ensemble densities in $P^{MC}$, in order to control statistical errors and computational cost.  
This mixed algorithm is elaborate, so the different approaches for each of the various components are described in detail in the following sections.

The material microstructure is evolved over time-step, $\tau$,  by operator splitting into five stages.
First, ME-MC reactions for rapidly evolving MC clusters (i.e., those with small $n_{v}$) are included  (Sec.~\ref{MEMCtext}) with a Markov chain method. 
Second, the ME-MC reactions for the large, slowly-evolving clusters are evaluated by Poisson-distributed random walks in composition space for each possible reaction with ME species (Sec.~\ref{MEMCtext}).  
Third, all MC-MC reactions are evaluated with the Markov Monte Carlo method. (Sec.~\ref{MCMCtext})
This completes the evolution of $P^{MC}$ over $\tau$.
The fourth stage integrates the ME including the  {\it average} source and sink terms from MC defects and dislocations (Sec~\ref{MEMEtext}). 
This completes the evolution for the void/bubble $P$. 
At this point, clusters may be exchanged between $P^{ME}$ and $P^{MC}$, without affecting the instantaneous total $P$ in any way (Sec.~\ref{ME2MCSec}).
This procedure may create new MC elements or eliminate existing ones, in order to control the growth of $N^{MC}$ versus time.
Fifth and finally, dislocation evolution is performed using a previously-described model \cite{WOLFER:1985}.

Overall numerical accuracy is monitored through the conservation of host and helium atoms.
Operator splitting of the evolution equations causes first-order  time-step errors.
However, conservation errors are dominated by differences between the ME and MC treatment of reactions in Sec.~\ref{MEMCtext} 
(i.e., continuous reactions versus discrete, stochastic events). 
These artifact statistical fluctuations are most important at low temperatures and especially during incubation, when $N^{MC}$ is smaller, defect annihilation dominates, and little net swelling occurs.
They must be carefully controlled, since the transient period represents a sort of barrier-crossing problem, 
with nucleation of stable voids and concomitant, self-consistent changes in the vacancy/interstitial populations as the barrier.
Any artificial Monte Carlo noise must not  spuriously affect the crossing 
into the steady-state.
In other words, $N^{MC}$-dependent fluctuations in the net vacancy content must not significantly promote or inhibit void nucleation.
In practice, 
stable cavities form naturally under the vacancy supersaturation and essentially irreversible aggregation of helium, and the volumetric swelling behavior is not unduly sensitive to $N^{MC}$ for the situations considered here.

\subsection{ME-ME reactions}
\label{MEMEtext}

Small defect clusters develop at high concentrations under irradiation, and so they dominate the system of  reactions.  
However, they quickly reach a quasi-stationary distribution wherein further reactions cause little change in their densities;  i.e.,  the majority of their reactions subsequently cancel one another.
It is much more efficient to treat the net reaction rates in a continuum approximation
rather than to explicitly account for individual reactions.
The ordinary differential equation solver, VODE,
provides an optimized treatment of stiff, nonlinear reaction equations \cite{VODE},
given  $f_n={{d\rho_n}\over{dt}}$ (Eqs.~\ref{MEQN} and~\ref{RATEEQN}) and the Jacobian, $J_{nm} = {{\partial f_n}\over{\partial \rho_m}}$ for all species.  
The computational cost increases rapidly with the number of coupled equations, hence the
cluster domain is limited to $0\le n_{vac}\le N_{vac}^{ME}$ and $0\le n_{hel} \le N_{hel}^{ME}$.
Typically, $N_{vac}^{ME}$ = 10-100 and $N_{hel}^{ME}$ = 2-10.
Some terms are excluded from the Master Equation so that all reaction products remain within 
this finite domain.
Clusters with $ 0\le n_{vac}\le N_{vac}^{ME}/2$ and $ 0\le n_{hel} \le N_{hel}^{ME}/2$ may undergo any mutual reactions, but no other ME clusters may undergo any reactions.
These latter clusters are frozen in size, so their density only increases as reaction products accumulate.  
Frozen clusters eventually transfer to the MC distribution as described in Section~\ref{ME2MCSec},
after which they will undergo reactions normally.

With reaction constraints and separate ME and MC distributions, 
the vacancy Eq.~\ref{MEQN} becomes:
\begin{align}
{{d\rho_v}\over{dt}} =& \enskip g_v(t)   +K({ v_2,i})\rho_{v_2}(t)\rho_i(t) \cr
&+ \sum_{\mathbf n\in ME} \biggl[- K({{\mathbf n},v})\rho_{\mathbf n}(t) \rho_v(t) 
+  K({{\mathbf n},0})\rho_{\mathbf n}(t)\biggr]
U\bigl({1\over2}{\mathbf N}^{ME}-{\mathbf n}\bigr)
\cr 
& -\biggl( \overline{S^{fast}_{v}}+S^{slow}_{v}(t_0)\biggr)\thinspace\rho_v(t)  + \biggl(\overline{S^{fast}_{0}}
+S^{slow}_{0}(t_0)\biggr)
\label{VTRUNC}
\end{align}
restricting the sums over $\mathbf n\in{ME}$ to reactive defects. 
Eq.~\ref{KEQN} also parametrizes unary vacancy emission reactions as the ${\mathbf n}$-null reaction, 
$K({{\mathbf n},0})=Z_{\mathbf n-v,v} A_{\mathbf n-v,v}D_{\mathbf n-v,v} c_v^{[\mathbf n-v]}$.
$S$ includes the external source and sink terms for reactive elements of $P^{ME}$; it accounts for ME reactions with defects in $P^{MC}$ and with dislocations.
Vacancy absorption at MC defects and dislocations is parametrized by $S_v$, and vacancy emission by  $S_0$.
The vacancy sinks and sources include separate terms that either evolve slowly or rapidly with time.  The coefficients are obtained in Sec.~\ref{MEMCtext}.
The rest of Eq.~\ref{MEQN}  takes similar form, with sinks $S_i$, $S_{vh}$, or $S_{h}$.  
(Only vacancies can be thermally emitted from defect clusters, so $S_0$ is the only source term.)

Operator  splitting over the time-step, $\tau$, is such that  external source and sink terms  $S$ are held constant as $P^{ME}$ evolves.
$S$ is divided into terms that evolve slowly or rapidly with time.  
The bar indicates an average of the sink strength over the time-step, from $t_0$ to $t_0+\tau$, 
useful for rapidly evolving MC clusters,
while slowly-evolving dislocations and large MC voids are simply evaluated at the beginning $t_0$  
(see also Sec.~\ref{MEMCtext} for further details).

The constrained coalescence Eq.~\ref{RATEEQN} becomes: 
\begin{align}
{{d\rho_{\mathbf n}}\over{dt}} = g_{\mathbf n}(t) 
&
  +\biggl \{ 
  \sum_{\mathbf m \in\{v,vh,h\}} K(\mathbf m,{\mathbf n}-{\mathbf m})  \rho_\mathbf m(t)  \rho_{{\mathbf n}-\mathbf m}(t) 
     \bigl (1-{\delta_{\mathbf m,\mathbf n-\mathbf m}\over2}\bigr)
  \biggl( 1-  U\bigl({\mathbf m}-\mathbf n\bigr ) \biggr) \cr
& \hskip 3truecm  \times U\biggl({1\over2}{\mathbf N}^{ME}-({\mathbf n}-\mathbf m)\biggr)
   U\biggl({1\over2}{\mathbf N}^{ME}-\mathbf m\biggr)
\cr
&
  -\sum_{{\mathbf m}\in\{v,vh,h,i\}} K({{\mathbf m},{\mathbf n}})\rho_{{\mathbf m}}(t) \rho_{\mathbf n}(t)  
 U\biggl({1\over2}{\mathbf N}^{ME}-{\mathbf n}\biggr)
\cr
&
  +\biggl[ K(i,{\mathbf n}+v)  \rho_i(t)  \rho_{{\mathbf n}+v}(t) 
  +K({{\mathbf n}+v,0})  \rho_{{\mathbf n}+v}(t) \biggr]   
    U\biggl({1\over2}{\mathbf N}^{ME}-({\mathbf n}+v)\biggr)
\cr
&
 -\biggl[  K(i,{\mathbf n})  \rho_i(t)  \rho_{{\mathbf n}}(t)
  + K({{\mathbf n},0})  \rho_{{\mathbf n}}(t) \biggr] 
  \enskip  U\bigl({\mathbf n}-v\bigr)
    U\biggl({1\over2}{\mathbf N}^{ME}-{\mathbf n}\biggr)
\cr
&+ \biggl\{
- \sum_{\mathbf m\not\in\{v,vh,h,i\}} K({\mathbf m},{\mathbf n}) \rho_{\mathbf m}(t) \rho_{\mathbf n}(t)   
 \enskip U\biggl( {{ {\mathbf N}^{ME} }\over2}-{\mathbf n}\biggr) 
 U\biggl( {{ {\mathbf N}^{ME} }\over2}-{\mathbf m}\biggr) 
 \cr
&
  +  {1\over2} {\sum_{\mathbf m\not\in \{v,vh,h,i\}}}^{\negthinspace\negthinspace\negthinspace\negthinspace\negthinspace\negthinspace\negthinspace \prime} K({\mathbf n}-{\mathbf m},{\mathbf m}) 
  \rho_{{\mathbf n}-{\mathbf m}}(t) \rho_{\mathbf m}(t)   \enskip 
  \biggl (1-U\bigl({\mathbf m}-\mathbf n\bigl) \biggr)
\cr
&  \hskip 3truecm \times U\biggl({{{\mathbf N}^{ME}}\over2}-({\mathbf n}-{\mathbf {\mathbf m}})\biggr)  
  U\biggl({{{\mathbf N}^{ME}}\over2}-{\mathbf m}\biggr) 
  \biggr\}
\cr
 &
 - \biggl(\overline{S^{fast}_{\mathbf n}}\thinspace +S^{slow}_{\mathbf n}(t_0) \biggr) \thinspace \rho_{\mathbf n}(t)
\label{TRUNC}
\end{align}
for clusters $\mathbf m, \mathbf n\in {\rm ME}$, and $\mathbf n \not\in\{v,vh,h,i\}$. 
The primed summation excludes $\mathbf n - \mathbf m \in \{v,vh,h,i\}$, since the monomer reactions are evaluated separately.  
$S$ includes any reactions of the $\mathbf n$-mer 
with the MC clusters and with dislocations.
There are no reactions that consume frozen clusters, so their concentration increases with time.

A subset of the disallowed reactions would produce clusters that  still lie within the ME domain.  
These have been excluded, for simplicity and to better resemble an earlier scheme for monomer aggregation \cite{Surh:2004}. 
Specifically, a homogeneous boundary condition is imposed on the Fokker-Planck equation in  Ref.~\cite{Surh:2004}, at $n=N_{vac}^{ME}$.
Clusters that grow to the boundary are removed from 
the Master Equation treatment and accumulated separately, during which time they are 
prevented from changing size.  
This is equivalent to keeping those $N_{vac}^{ME}$-sized clusters 
within $P^{ME}$ but disabling all of their reactions.  
Frozen clusters are then intermittently transferred to $P^{MC}$, where they are no longer constrained \cite{Surh:2004}.

Ideally, the ME domain will encompass all non-zero generation terms, $g_{\mathbf n}$, 
and include as many sub-critical or transient defect cluster species as possible.
A relatively small domain of $N_{v}^{ME} \simeq 60$,  $N_{h}^{ME} \simeq 4$ is chosen here, reflecting the computational demands that coalescence imposes.  
Similarly to \cite{Surh:2004}, the solution is recorded at exponentially-increasing intervals.
This time-step is irrelevant to the ME evolution itself, which advances by adaptive sub-steps.  
However, $\tau$ controls errors from operator splitting of the evolution equations, and it governs the creation of MC elements, as described below.  

Because the sink/source terms, $S$, are evaluated by a discrete MC method, they introduce a fictitious noise to the continuum rate equations.  
This partly manifests as step-function discontinuities in the sink strength over successive time-steps,
which in turn causes transient relaxation in the concentrations of the ME species.
The numerical solution tries to accurately follow the transients, potentially making the fully coupled, non-linear evolution inefficient, when large time-steps are otherwise possible.
Rather than faithfully simulating these spurious transients at late times, it may be preferable to solve the monomer concentrations (Eqs.~\ref{MEQN}, ~\ref{VTRUNC}, etc.) in the quasi-stationary approximation 
after any real transient behavior (due to the abrupt onset of irradiation or other changes in environmental parameters) has concluded. 
Eq.~\ref{TRUNC} for dimers and larger clusters may then be solved while holding the monomer concentrations fixed over the time-step.   
In practice, after a brief transient, the results are comparable to those obtained from the full, coupled, non-linear ME solution.

\subsection{Transfer between ME and MC domains}
\label{ME2MCSec}

A majority of the ME elements  in a small multi-dimensional 
domain will lie near its boundary, and so the majority of the ME cluster species 
will be artificially frozen.  
The constraints on the defect clusters are only lifted after they are transferred to $P^{MC}$.
There are three desiderata to this transfer process.
Foremost, it must minimize any systematic, constraint-induced errors,
therefore the density of frozen clusters must be small compared to the rest of $P$.
Secondly, the MC computational cost must be controlled, therefore $N^{MC}$ must be kept small.
Rather than increasing $N^{MC}$ at every opportunity, frozen clusters at ${\mathbf n}\in{\rm ME}$ are allowed to accumulate until exceeding a spawning threshold density, $\rho_{\mathbf n}^{ME} > \rho_{sp}$, as in \cite{Surh:2004}. 
At the end of that time-step, a portion of the accumulated density is removed from $P^{ME}$ and transferred to a new element of $P^{MC}$, incrementing $N^{MC}$.
\begin{equation}
({\mathbf n},  \rho_{{\mathbf n}}  )^{ME} \rightarrow ({\mathbf n},  \rho_{{\mathbf n}} -\delta\rho)^{ME}
+ ({\mathbf n}; {{\delta\rho}})   ^{MC}
\end{equation}
with the ME and MC compositions coinciding.  
If the accumulated $\rho_{\mathbf n}^{ME}  > \rho_{sp}$ after each time-step, 
then the accumulating clusters are effectively never constrained. 
Finally, it is imperative to minimize any $N^{MC}$-dependent Monte Carlo statistical error.
Individual MC elements with the largest $\rho$ will contribute the most to this error.
Therefore, if $\rho_{{\mathbf n}}^{ME}  \gg \rho_{sp}$ at the end of a time-step,  
then $\Delta N > 1$ new MC elements are created, as:
\begin{equation}
({\mathbf n},  \rho_{{\mathbf n}}  ) ^{ME} \rightarrow  ({\mathbf n},  \rho_{{\mathbf n}} -\delta\rho)^{ME}
+ \Delta N  \times({\mathbf n}; {{\delta\rho}\over{\Delta N}})   ^{MC}
\end{equation}
Equivalently, MC elements  with large $\rho$ may be split into identical macroparticles with smaller densities.
The chosen values for $\tau$, $\rho_{sp}$, and the functional dependence of $\Delta N$ on $\delta \rho$ control the $N^{MC}$-related statistical error and computational cost for a simulation.
Typically, ${\rm log}_2(\Delta N)= {\rm Int}({\rm log_{30}}(\delta\rho/\rho_{sp}))$.

For example, the distribution in Fig.~\ref{Fig2} shows the production of many MC macroparticles containing 2-4 helium atoms; these react and form a plume that extends to $n_{v}\simeq100$. 
The ME domain used in this example also includes frozen cluster species with 5-9 helium; 
these species  have not yet reached the threshold density.  
They eventually spawn  MC elements, but
at a much slower rate than for the near-critical sizes of 2-4 helium.
Even at this early time, the total density of constrained ME clusters is 
small compared to the MC population so constraint errors are minimized.

Since $\rho_{sp}$ cannot be made arbitrarily small in practice, it is useful to add a second transfer mechanism.  
When a pre-existing MC element at $\mathbf n$ falls inside the 
frozen ME domain, the change:
\begin{equation}
({\mathbf n},\rho_{{\mathbf n}} )^{ME}+({\mathbf n},\rho_{{\mathbf n}} ) ^{MC}
\rightarrow
({\mathbf n},\rho_{{\mathbf n}} -\delta\rho)^{ME}+({\mathbf n},\rho_{{\mathbf n}} +\delta\rho)^{MC}
\end{equation}
leaves the total distribution unchanged.
$N^{MC}$ remains constant, so the calculation remains tractable.
In practice, the maximum amount $\delta\rho\leq \rho_{\mathbf n}$ is transferred until the receiving MC element reaches a cutoff density, $\rho_{\mathbf n}^{MC}+\delta \rho \leq \rho_{max}$  (where typically, $\rho_{max}\simeq 2\rho_{sp}$ to $10\rho_{sp}$.)
The cutoff prevents over-weighting of individual Monte Carlo elements so as to control the statistical error.  

At low temperatures, a very high density of small bubbles can coexist with a moderate density of large, low-pressure voids.
Such distributions are most efficiently treated by making $\rho_{max}$ size-dependent, so that the maximum macroparticle densities are high in the region of bubbles, but low in the region of voids.
Macroparticles can freely wander between the two regions.
Accordingly, if  macroparticle $A$ moves to a region where $\rho_A>\rho_{max}$, it may be split into two identical parts;
or if two MC elements at the same coordinate have $\rho_A+\rho_B < \rho_{max}$, they may be united into one.

In problems of reversible nucleation and growth, small MC clusters may shrink and disappear.
It is computationally inefficient to follow unstable clusters by Monte Carlo methods.  
Accordingly, macroparticles of the smallest vacancy clusters 
(with both $n_{vac}<N_{vac}^{min}<N_{vac}^{ME}$ and $n_{hel}=0$) are deleted
at the end of each time-step and their density returned to the corresponding element of $P_{ME}$.
(The numerical solution of the ME automatically accommodates any subsequent transients by adjusting its internal time-steps.)  
The minimum MC size  should be large enough that macroparticles at the threshold 
only rarely shrink to monomer sizes during the interval $\tau$.  
It should also be far enough from $N_{vac}^{ME}/2$  that the cycle ME$\rightarrow$MC$\rightarrow$ME (involving creation of a new macroparticle, shrinkage of the constituent clusters, and transfer of that element back to $P^{ME}$) is infrequent.  
In practice, ${\mathbf N}^{min}={\mathbf N}^{ME}/4$ is used, and these two criteria are accomodated by taking the largest possible ${\mathbf  N}^{ME}$.  
Helium clusters are never returned from MC to ME distribution; helium emission is not permitted, so the clusters will only grow along the helium axis.

In the examples considered here, all ME clusters are sub-critical for $N_{vac}^{ME}\simeq60$, 
so that newly-created MC particles frequently shrink and are annihilated.  
This is especially true at low temperatures, when the proliferation of small voids favors vacancy/interstitial recombination.
Here, this "rare event problem" for nucleation of stable voids from small vacancy clusters is at least
improved from conventional kinetic Monte Carlo methods, where even the monomers would be treated stochastically.
Ultimately,  direct application shows this mixed scheme is suitable for radiation damage calculations to high doses.  

\subsection{MC-MC reactions}
\label{MCMCtext}

Coalescence problems 
are frequently treated  by a Markov Monte Carlo method \cite{GILLESPIE:1972}.   
A straightforward approach defines a finite volume, $V$, containing $N$ (i.e., $N^{MC}$) discrete clusters of sizes $\{\mathbf n\}$ that stochastically evolve to a new $N - 1$ population $\{{\mathbf n}^\prime\}$ through the binary coalescence of any pair of particles.  
The average rate of reaction between the $i$th and $j$th particles is 
simply $K({\mathbf n_i,\mathbf n_j})/V^2$ per unit volume.
The total rate of reaction of all $N$ clusters is $R_{N}$, where 
\begin{equation}
R_i = \sum_{k=1}^{i} R_{k,N}
\label{RVECTOR}
\end{equation}
and
\begin{equation}
R_{i,j} = \sum_{k=1}^{j} {1\over2} K({\mathbf n_i,\mathbf n_k})/V
\label{RROWS}
\end{equation} 
in terms of the sum over reactions in the entire volume, $V$,
assuming they are uncorrelated and occur in parallel.   
$R_{i,N}$ is proportional to the rate at which cluster $i$ reacts with all other clusters.

A stochastic sequence of discrete reactions may be constructed from these parameters.   
The random interval, $\xi$, to the next reaction is obtained from a uniform variate, $x\in(0,1]$, as \cite{MACKEOWN}:  
\begin{equation}
\xi = -{\rm ln}(x)/R_{N}
\end{equation}
The first cluster of the random reaction pair, $i$, is selected with a probability proportional to $R_{i,N}$, from $y\in(0,1]$ and  
\begin{equation}
{R_{i-1}\over{R_{N}}}  < y \leq {{R_i}\over{R_{N}}}
\label{ISELECT}
\end{equation}  
where $R_{0} \equiv 0$.
Finally, the reaction counterpart, $j$, is identified from  $z\in(0,1]$ 
and 
\begin{equation}
{{R_{i,j-1}}\over{R_{i,N}}} < z \leq {{R_{i,j}}\over{R_{i,N}}}
\label{JSELECT}
\end{equation}
 with $R_{i,0}\equiv 0$.
This selects $j$ with a probability proportional to ${1\over2}K({\mathbf n_i,\mathbf n_j})/V$.   
The procedure repeatedly selects new $x$, $y$, and $z$ for the next event, increments the system time by $\xi$, performs the reaction $i+j$, and recalculates $R$ for the next iteration.  This repeats until the elapsed time exceeds $\tau$.  Since the last reaction falls outside the desired interval, it is discarded without being performed.  The procedure may then be started anew for the next time-step.  

The choice of two random numbers to select the pair, $i, j$, differs from the usual scheme, where the pair is selected from a single value.  
In either case, the search for $i$ and $j$ takes $o(log_2(N))$ operations using the method of bisection
 \cite{NUMREC3}.  
However, separate selection of $i$ and $j$ makes it possible to record all $R_{m}$ with $o(N)$  storage space and a one-time computational effort of $o({N}^2 )$.  
Once $i$ is determined, $R_{i,m}$ may be tabulated with $o(N)$  effort for all $m$, so the full matrix need not be stored.  
Finally, after $i$ and $j$ react, the $R_m$ may be updated with $o(N)$ effort by replacing only those terms involving the old clusters $i$ and $j$ with the results for a single new, coalesced cluster, and re-indexing to account for the lost cluster.
Since $R_{N}$ is an extensive quantity for a given total density, 
evolution of $N$ particles over a finite interval requires $o({N}^2 )$ effort and 
$o(N)$ storage.
Specifying the binary reaction rate coefficients, $K$, as a half-triangular matrix increases the efficiency marginally.

This MC scheme has difficulty modeling widely varying concentrations of reactants (e.g., the monomer density is typically orders of magnitude higher than the large clusters for radiation damage). 
Also, $N$ decreases after every coalescence, which increases the statistical noise.
There are methods that preserve $N$ 
\cite{SMITH:1998,MUKHERJEE:2003}, but it is  possible to encompass a wider range of densities
at the same time.
In the approach taken here, the discrete MC elements are macroparticles, widely used in, e.g., non-equilibrium simulations of plasma physics \cite{MACRO}.  
(This is distinct from related, "weighted particle" schemes for coagulation  \cite{BABOVSKY:1999,EIBECK:2001,SABLEFELD:1996}.)
Here, the $j$-th macroparticle in the system consists of an ensemble of clusters all of the same composition $(\mathbf n_j,\rho_j)$.  
Consistent with the Gillespie procedure, macroparticle reactions are evaluated discretely, so clusters in an ensemble react simultaneously but otherwise stochastically.  
However, here reactants will generally have different ensemble densities, $\rho_L < \rho_H$, which are independent of their sizes/compositions, $\mathbf n_L,\mathbf n_H$.  The lower-density macroparticle, $L$, reacts completely, leaving behind an unchanged portion $\rho_H-\rho_L$ of clusters from the higher-density ensemble, $H$.  The total cluster density declines, but $N$ stays constant, and $N$-dependent errors remain steady over time.

Macroparticle reaction rates (analogous to Eq.~\ref{RROWS}) are defined so as to reproduce the continuum limit as $N\rightarrow\infty$. 
Pairs $i$ and $j$, with $\rho_i<\rho_j$, react according to:
\begin{equation}
\begin{alignedat}{5}
&(\mathbf n_i,  \rho_i) + (\mathbf n_j, \rho_j) 
    \rightarrow 
   (\mathbf n_i + \mathbf n_j, \rho_i)  +   (\mathbf n_j, \rho_j-\rho_i)  
\label{MCMC1}
\end{alignedat}
\end{equation}
at an average rate of $K({\mathbf n_i,\mathbf n_j}) \rho_j $.
Two macroparticles of the same density ($ \rho_i =\rho_j=\rho;\enskip i \neq j$) react as:
\begin{equation}
\begin{alignedat}{5}
&(\mathbf n_i,\rho)    + (\mathbf n_j,\rho)  
   \rightarrow 
   (\mathbf n_i+\mathbf n_j,\rho/2) + (\mathbf n_i+\mathbf n_j,\rho/2)    
\end{alignedat}
\end{equation}
at an average rate of $K ({\mathbf n_i,\mathbf n_j}) \rho$.
The product is simply split into two equal pieces so that $N$ remains constant.  
Finally, the individual clusters within a single macroparticle ensemble may coalesce with one another, so there is also a unary reaction process:
\begin{equation}
\begin{alignedat}{5}
&(\mathbf n_i,\rho_i) \rightarrow (2 \mathbf n_i,\rho_i/2)  \quad       
\end{alignedat}
\end{equation}
which also proceeds at an average rate of $K({\mathbf n_i,\mathbf n_i}) \rho_i$.
This possibility modifies Eq.~\ref{RROWS} to include a non-zero reaction rate for $i=j$.  

Macroparticle dynamics 
never corresponds to an atomistic simulation for finite $N$.  
Instead, this is a distinct, approximative discretization of the continuum equations themselves, in the same spirit as earlier approaches \cite{Surh:2004}.  
Again, $P(t)$ is approximated here by a sparse set of elements without arbitrarily imposing some coarse-graining of finite difference equations for the distribution.  
Since the computational cost scales as $o(N^2)$ for a dense reaction matrix, the method is also efficient.
This is especially advantageous in higher dimensions, e.g., in describing helium-vacancy-impurity clusters.

\subsection{ME-MC reactions}
\label{MEMCtext}

Additional schemes are required for treating reactions between ME and MC 
elements. 
In the continuum approximation, reaction with external entities, $\mathbf n \not\in $ ME,  introduces 
unary sink terms to the rate equation for  $\mathbf m \in{\rm ME}$, cf. Eqs.~\ref{VTRUNC},~\ref{TRUNC}: 
\begin{equation}
S_{\mathbf m}(t) \rho_{\mathbf m}(t)=  \biggl[\sum_{\mathbf n\in MC}K\bigl(\mathbf m,\mathbf n(t)\bigr) \rho_{\mathbf n}(t) 
+K(\mathbf m,d)\rho_d(t)
\biggr]\rho_{\mathbf m}(t)
U(\mathbf N^{ME}/2 - \mathbf m)
\label{MEMC1}
\end{equation}  
where the summation includes all elements $\{(\mathbf n(t),\rho_{\mathbf n}(t)\}^{MC}$ at time $t$ and where $K(\mathbf m,d)$ includes reactions with network dislocations.
The sink  term, $S$, is identically zero for constrained ME defects.
At present, $K(\mathbf m,d)$ is only nonzero for $\mathbf m=(1,0), (-1,0)$ and for vacancy emission $K(\mathbf 0,d)$.

The counterpart to Eq.~\ref{MEMC1} is expressed for $\mathbf n\in{\rm MC}$ in the macroparticle scheme by:
\begin{equation}
\begin{alignedat}{5}
&({\mathbf n},\rho_{\mathbf n})^{MC}\rightarrow (\mathbf m+\mathbf n,\rho_{\mathbf n})^{MC}
\label{MEMC2}
\end{alignedat}
\end{equation}
as a discrete reaction with an average rate of $K(\mathbf m,\mathbf n) \rho_{\mathbf m}$.
A stochastic sequence of reactions at these average rates will approach 
the continuum behavior of Eq.~\ref{MEMC1} in the limit $N^{MC}\rightarrow\infty$.
A single reaction can change a macroparticle size, cross-section, and reaction rate substantially, 
if $\mathbf m$ is comparable in size to $\mathbf n$.  
Accordingly, ME-MC reactions for such "small" MC clusters are included by the Markov Monte Carlo scheme described above, and the reaction parameters are immediately updated to reflect the change, 
before evaluating the next reaction.

Reaction events are randomly performed from the  $N^{MC}\times N^{ME}$ matrix of reaction rates, at overall rate $Q$.  
If the next event occurs within the desired interval, the $i$th MC element is selected as a reactant with probability $Q_i/Q$, where:
\begin{equation}
Q_{i} =   \sum_j K(\mathbf n_i,\mathbf m_j)\rho_{\mathbf m_j}
\label{QVECTOR}
\end{equation} 
for reactive elements $\mathbf m_j\in ME$ and:
\begin{equation}
Q = \sum_{i=1}^{N^{MC}} Q_i
\end{equation}
The $j$th ME element is selected as a reactant with probability $K(\mathbf n_i,\mathbf m_j)\rho_{\mathbf m_j}/Q_i$.
Finally, the time index is updated, the reaction is performed, and $Q$ is revised.
This is analogous to the Markov procedure for MCMC reactions, except that the reaction matrix is full-rectangular rather than half-triangular and that the rates are always proportional to the density of the ME reactant.

As for the corresponding evolution of $P^{ME}$, the instantaneous source/sink terms, Eq.~\ref{MEMC1}, change after each discrete reaction event in $P^{MC}$, 
possibly multiple times during the interval $\tau$.
It is not computationally practical to evolve $P_{ME}$ over each individual Markov sub-step, $\xi$, to account for this.
Instead, $P_{ME}$ is evolved over the full time-step $\tau$ by operator splitting, after all ME-MC and 
MC-MC reactions in $P_{MC}$ are performed. 
To minimize any convergence error, the instantaneous sink strength can be replaced with a weighted time average over the interval:
\begin{eqnarray}
\overline{S^{fast}_{\mathbf m}} &=& {1\over\tau} \int_{t_0}^{t_0+\tau} dt \bigg\lbrack \sum_{\mathbf n\in MC}K(\mathbf m,\mathbf n(t)) \rho_{\mathbf m}(t)\bigg\rbrack \\
&=& {1\over\tau} \sum_j \xi_j \bigg\lbrack \sum_{\mathbf n}K(\mathbf m,\mathbf n(t_{j-1})) \rho_{\mathbf m}(t_{j-1})\bigg\rbrack
\label{SPRIME} 
\end{eqnarray}
finally expressed as a sum over sub-intervals, $\xi_j$, between discrete reaction times, $t_j$.

Such attention to detail is unnecessary for large MC clusters (and for network dislocations), where rapid reactions with highly mobile defects (i.e., small $\mathbf m $) do not  substantially change the sink strength over short intervals.  
Thus, it is sufficient to update parameters for the large $\mathbf n$ clusters at the end of each time-step.
In this case, MC clusters are evolved using a Poisson-distributed random variate, $P(x)$, \cite{GILLESPIE:2000,NUMREC5} for the number of reactions that occurs during $\tau$.  
These MC elements are only updated at $t_0+\tau$, with all reactions accumulated in each of the $N_{ME}$ channels:
\begin{equation}
(\mathbf n,\rho_{\mathbf n}) \rightarrow \bigg(\mathbf n + \sum_{\mathbf m\in ME} \mathbf m P\big\lbrack \tau K({\mathbf m,\mathbf n})  \rho_{\mathbf m}\big\rbrack, \enskip \rho_{\mathbf n}\bigg)
\label{MCPOISSON}
\end{equation} 
Equation~\ref{MCPOISSON} is the discrete analogue of the Gaussian-distributed random walk used previously \cite{Surh:2004}.
The corresponding ME sink term is:
\begin{equation}
S^{slow}_{\mathbf m}(t_0) =  \sum_{\mathbf n \in MC}K\big(\mathbf m,\mathbf n(t_0)\big) \rho_{\mathbf n}(t_0) + K(\mathbf m,d)\rho_d(t_0)
\label{SPOISSON}
\end{equation}  
including the dislocation contribution, assuming that $\rho_d(t)$ is slowly changing.

Finally,  discrete reactions could also be evaluated by a rejection method, given a Majorant kernel $M(\mathbf m,\mathbf n)\ge  K(\mathbf m,\mathbf n)$ \cite{SABLEFELD:1996}.  
For example, the reaction rates, $M$,  can be evaluated on a coarse grid of ${\mathbf n}_i$ and all 
reactants ${\mathbf n}_i\le{\mathbf n}\leq{\mathbf n}_i+1$ be treated alike.
In another approach, $M$ may be chosen to be a sum of products \cite{EIBECK:2001}, 
\begin{eqnarray}
M(\mathbf m,\mathbf n)  & = \enskip \vec {\cal M}(\mathbf m) \cdot \vec {\cal M}(\mathbf n)
\label{P1}
\end{eqnarray}
 It is then only required to evaluate $N^{MC}$ vectors, $\cal M$, (of one or more dimensions)  and to take dot products.
Either approach is easier than directly computing $N^{MC}(N^{MC}+1)/2$ binary rate coefficients for Eqs.~\ref{RVECTOR}, \ref{RROWS}.
The Majorant kernel is selected to be easy to evaluate and to predict a faster (or equal) event rate than the true system.
To correct for any overestimate, the time index is updated according to the usual Markov Monte Carlo procedure, 
but the reaction is only performed if
a uniform variate, $w\in(0,1]$ also satisfies $w\le K(\mathbf m,\mathbf n)/M(\mathbf m,\mathbf n)$.
Thus, excess events predicted by $M$ are rejected (with the required probability $1-K/M$).  
At present, the full reaction rate coefficients from Eq.~\ref{KEQN} can be evaluated very efficiently,
so this method is not employed here. 
However, it is expected to be 
advantageous when biased cavity-cavity, cavity-loop, and loop-loop reactions are included in the future.

\section{Results}

\subsection{Monomer aggregation model}

A high density of trapping/recombination centers is believed to delay the onset of  void swelling \cite{NELSON:1966,MANSUR:1986,MANSUR:1990}.  
Traps hinder void diffusion and coalescence and prolong the incubation stage.
The simplest trapping model assumes that all dimers and larger clusters are immobile: $D_{\mathbf n}\equiv0$ for all ${\mathbf n}\not\in\{v,vh,h,i\}$, so that the last two summations in Eq.~\ref{RATEEQN} are zero. 
If Eq.~\ref{MEQN} is solved separately from the remainder of the Master Equation (Eq.~\ref{RATEEQN}) in a quasi-stationary approximation, then that problem may be solved by existing 
methods \cite{Surh:2004,WEHNER:1985}.  
However, here the problem is simply treated as a limit case of Smoluchowski's coagulation equation.

\begin{figure}[h!]
\begin{center}
\includegraphics[width=12cm]{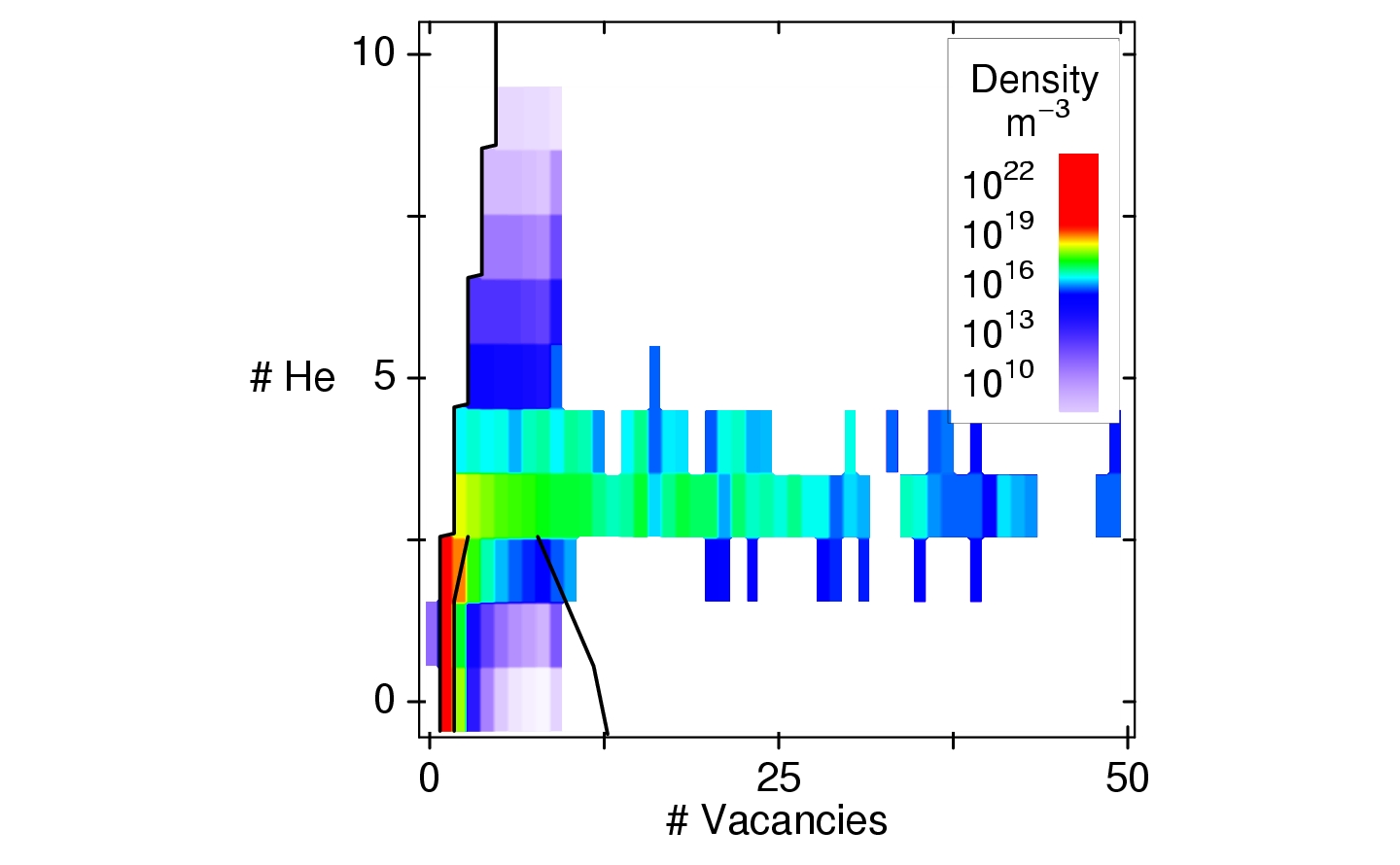}
\caption{A portion of the void/bubble distribution for a model with mobile monomer defects and sessile clusters, $\gamma=0.8 \gamma_0(T)$, at T=300 C, $10^{-6}$ dpa/s, and $32\times10^{-3}$ dpa.  
The largest void in the distribution contains 110 vacancies. The solid lines display the loci of stable and unstable equilibrium cluster compositions, based on average vacancy accumulation rates. 
This distribution has not been smoothed - the pixellated appearance reflects discrete cluster compositions.
(Fig2)}
\label{Fig2}
\end{center}
\end{figure}

\begin{figure}[h!]
\begin{center}
\includegraphics[width=12cm]{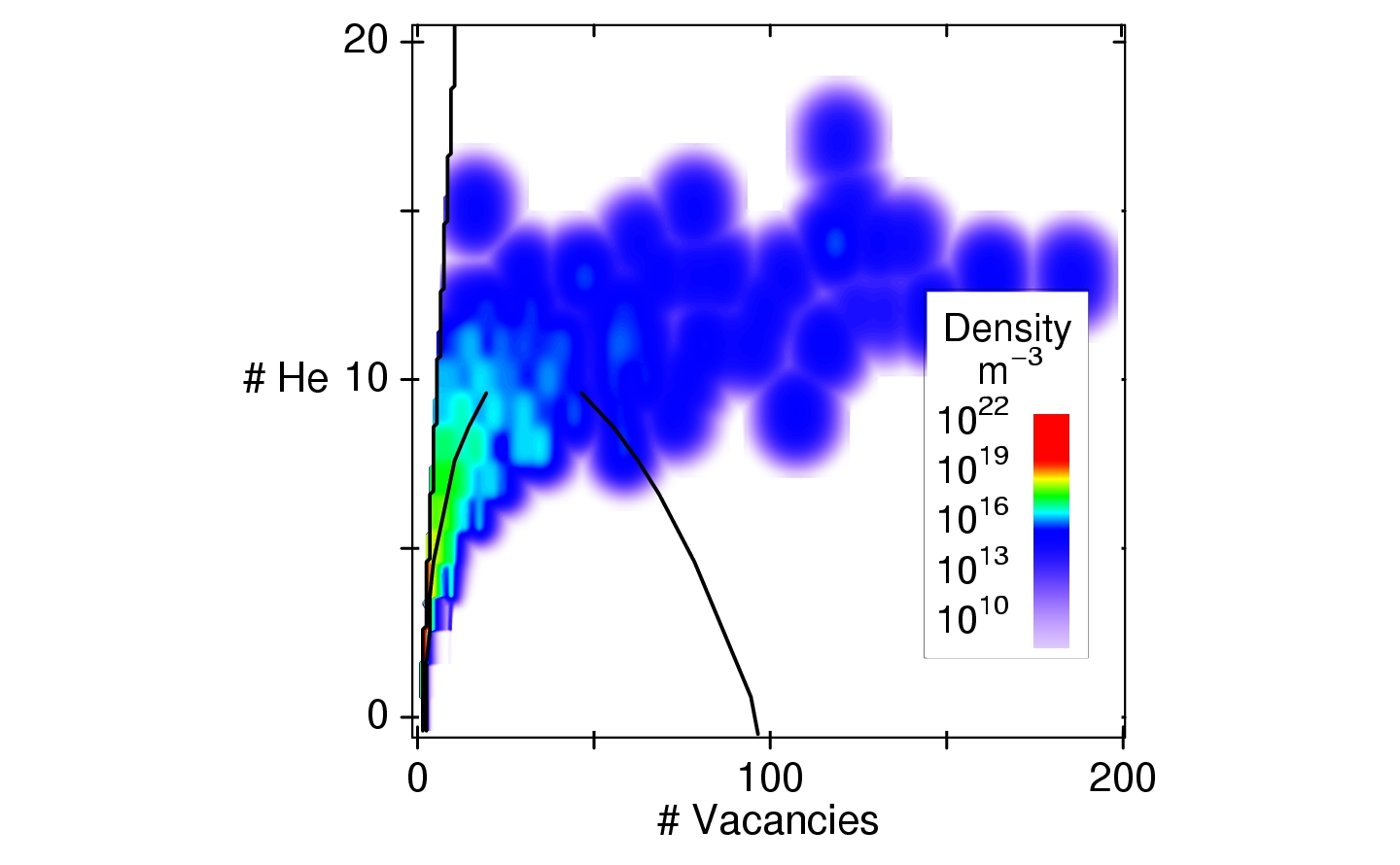}
\caption{A portion of the void/bubble distribution as in Fig.~\ref{Fig2}, but at T=500 C, $10^{-6}$ dpa/s, and $16\times10^{-3}$ dpa.    The distribution has been smoothed for the large clusters, where Monte Carlo data is increasingly sparse. The solid lines display the stable and unstable equilibrium cluster compositions.
}
\label{Fig3}
\end{center}
\end{figure}

\begin{figure}[h!]
\begin{center}
\includegraphics[width=12cm]{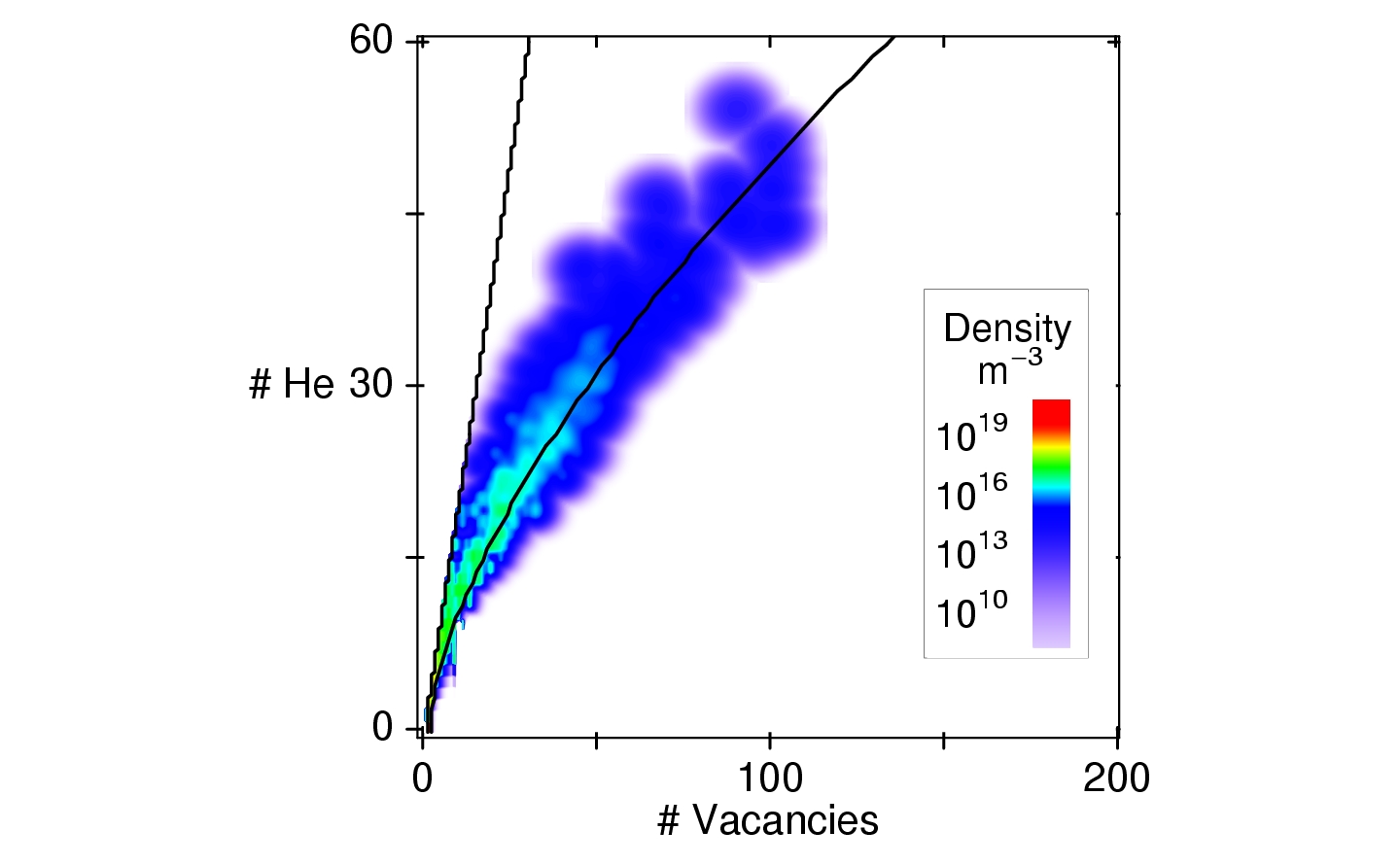}
\caption{The full void/bubble distribution as in Fig.~\ref{Fig3}, but at T=700 C.  The curved solid line locates the stable equilibrium bubbles; the critical void size is not visible on this scale.}
\label{Fig4}
\end{center}
\end{figure}

Initial cluster populations are shown in Figs.~\ref{Fig2}-\ref{Fig4} for  type-316 stainless steel  irradiated to low doses at $10^{-6}$ dpa/s and 300, 500, and 700 C.  
It is well-known that helium-vacancy clusters may be separated into distinct species (of equilibrium bubbles and stable or unstable voids), according to their size-dependent free energies.
Accordingly, the figures are marked with black lines where the net {\it average} vacancy addition rate for the defect clusters approaches zero.  
The leftmost black line in Figs.~\ref {Fig2}-\ref{Fig4} represents a hard wall for over-pressurized bubbles:  by fiat, bubbles cannot reach densities greater than 2 helium per vacant site.  
Here, this is imposed by disallowing further reactions with helium- and self-interstitials.
Other lines separate clusters that add or lose vacancies on average.
Small, over-pressurized bubbles tend to add vacancies until reaching the next line in the Figures, where stable helium bubbles are in dynamic equilibrium with the vacancy and interstitial population.  
(This approximates the line of bubbles with $P\simeq \gamma/2r$, which would be in equilibrium in the absence of a vacancy and interstitial supersaturation.)
The stability of these bubbles is reflected by their elevated concentration in that region, 
especially visible in Fig.~\ref{Fig4}.
Stable bubbles  tend to grow along the equilibrium line as they accumulate helium, 
while adjusting their vacancy content {\it on average} to remain in equilibrium.
Finally,  bubbles cannot exceed some critical helium content - larger clusters are stable voids that tend to add vacancies and grow to arbitrary size.
This is seen in Fig.~\ref{Fig3}; there the clusters grow along the line of stable bubbles until reaching a critical helium content (11 heliums), at which point they  grow by adding vacancies in excess of helium, forming a plume of rapidly-growing voids in the size distribution.

Voids are here simply taken to be cavities with higher vacancy/helium ratio than any bubble species of the same helium content.
An approximately parabolic region under the black curves bounds a set of small, unstable voids that tend to lose vacancies and shrink back towards the equilibrium bubble configuration.  
For example, this ranges from the origin to vacancy/helium compositions of (19,11) and (94,0)  in Fig.~\ref{Fig3}.  
The rightmost solid line identifies the critical or unstable equilibrium void compositions; 
larger voids tend to add net vacancies with time.
Note that a percentage of equilibrium bubbles in Fig.~\ref{Fig3} are able to fluctuate in vacancy content across the barrier of unstable voids.
That is, they become stable voids without having first reached the critical helium content.  
Similarly, helium dimers are readily able to cross the barrier of unstable voids in Fig.~\ref{Fig2}.
Very large voids ultimately become neutral (unbiased) sinks,
adding helium/vacancies at an average rate of 1:200 (based on anticipated asymptotic swelling of 1$\%$/dpa and model helium generation around 50 appm/dpa).  
Thus, voids approach a line of constant composition.

Except for a brief transient at the onset of irradiation, the vacancy monomer concentration decreases monotonically with time as the total sink strength of the microstructure rises with dose. 
After a few dpa, production of $\alpha$-particles also peaks, and the helium monomer concentration also declines.
During this extended period, equilibrium bubbles continue to grow by adding helium, they continue to reach the critical size, and they continue to become voids.
However, the critical size for equilibrium bubbles increases with time (as a function of declining $\rho_v$), and the rate of formation of new helium dimer nuclei and bubble growth rates decrease (as a function of declining $\rho_h+\rho_{hv}$).
This causes the rate of void formation to decrease gradually with time, giving
a broad void size distribution.
Eventually, the larger stable bubbles become TEM-visible, 
and  the overall size distribution becomes bimodal.

\begin{figure}[h!]
\begin{center}
\includegraphics[width=12cm]{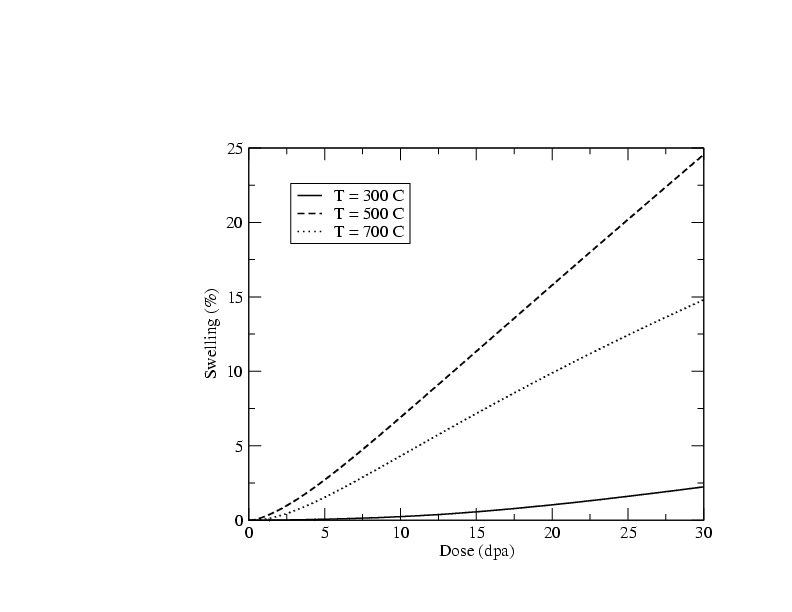}
\caption{Volumetric swelling curves versus dose in the model that excludes void-void coalescence.
The cavity surface energy is fixed at $\gamma(T)=0.4\gamma_0(T)$}
\label{NoCoalSwell}
\end{center}
\end{figure}

The time-dependent volumetric swelling for this model is shown at a series of temperatures in Fig.~\ref{NoCoalSwell}.
The low temperature system is initially dominated by large numbers of transient, unstable vacancy clusters (Fig.~\ref{Fig2}) that serve as recombination centers and suppress swelling.  
So many defect centers form that helium/vacancy ratios are kept low, and helium plays a reduced role in the initial evolution. 
As a result, the visible cavity density ($r>0.5$ nm) is sensitive to the surface energy parameter, $\gamma$: $\rho_{vis} = 5\times10^{23}$ m$^{-3}$ for $\gamma(T)=0.8\gamma_0(T)$ and $1\times10^{24}$ m$^{-3}$ for $\gamma(T)=0.5\gamma_0(T)$.  
Eventually, some vacancy clusters acquire significant amounts of helium, and the system is filled with  a high concentration of small equilibrium bubbles.
These function as recombination centers; they keep the vacancy supersaturation low so that few, if any, bubbles grow into stable voids.  
They also keep the asymptotic swelling rate small.
At and above 500 C, swelling is more a matter of helium bubble formation and growth towards critical sizes (Figs.~\ref{Fig3} and \ref{Fig4}).
The cavity density and swelling rates are therefore insensitive to $\gamma$.
The steady swelling rate of 0.8 \%/dpa at 500 C is consistent with void swelling in austenitic stainless steel \cite{Surh:2005,Surh:ERR}.  
At higher temperatures, the increased helium mobility results in fewer cavities (7-8$\times10^{20}$ m$^{-3}$ at 700 C), and a smaller density of bubbles  escape to become stable voids and contribute to steady swelling.

\subsection{Cluster coalescence model}

\begin{figure}[h!]
\begin{center}
\includegraphics[width=12cm]{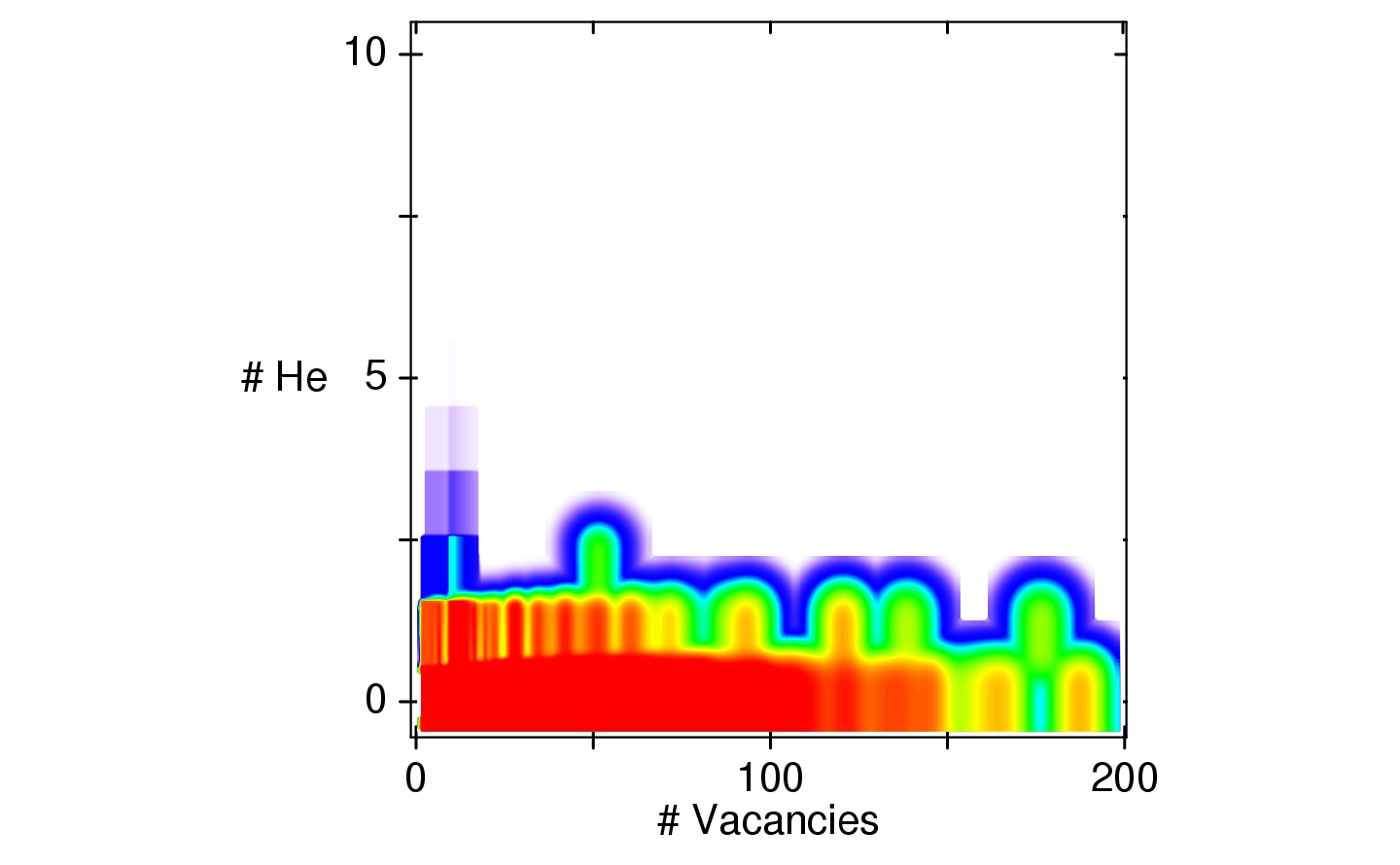}
\caption{A portion of the void/bubble distribution as in Fig.~\ref{Fig2} (300 C), but including void coalescence and with $\gamma(T) = 0.5\gamma_0(T)$.    The distribution has been smoothed for the large clusters, where Monte Carlo data is sparse. }
\label{Fig7}
\end{center}
\end{figure}

\begin{figure}[h!]
\begin{center}
\includegraphics[width=12cm]{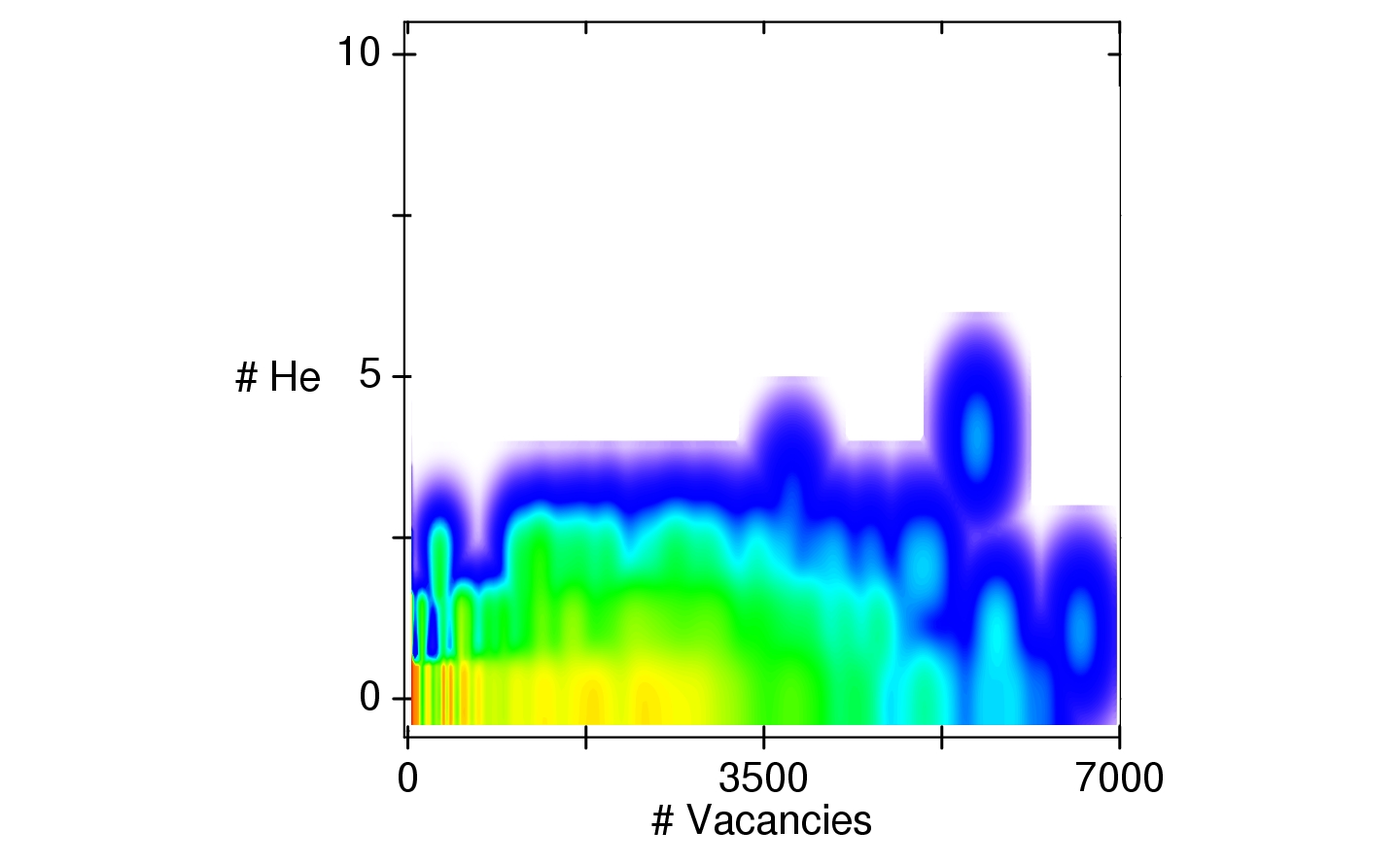}
\caption{The full void/bubble distribution as in Fig.~\ref{Fig4} (500 C), but including void-void coalescence and with $\gamma(T) = 0.5\gamma_0(T)$.}
\label{Fig8}
\end{center}
\end{figure}

\begin{figure}[h!]
\begin{center}
\includegraphics[width=12cm]{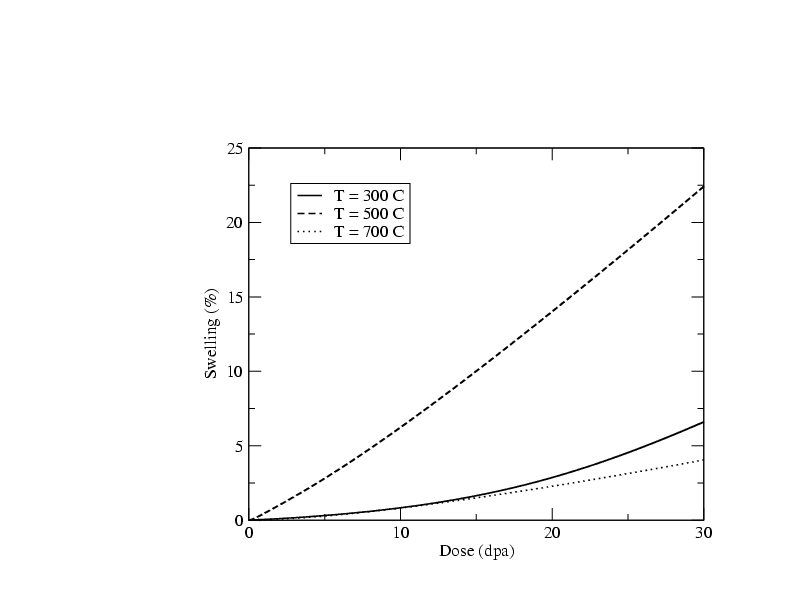}
\caption{Volumetric swelling curves versus dose in the model that includes void-void coalescence.
The cavity surface energy is set to $\gamma(T)=0.4\gamma_0(T)$.}
\label{SwellCoal}
\end{center}
\end{figure}

The other simplification of defect trapping is to neglect it entirely and 
assume that clusters diffuse freely according to their size.
The predicted void size distribution changes significantly
when coalescence is included.
This is seen in Figs.~\ref{Fig7} and \ref{Fig8}, 
for the same temperatures as in Figs.~\ref{Fig2} and \ref{Fig3}.
Coalescence reactions continually, preferentially consume the smaller, more mobile clusters.
The largest voids grow an order of magnitude larger through coalescence, making the distribution of stable void sizes substantially broader than before. 
Very large voids achieve such low diffusivities as to be effectively immobile;
this results again in a terminal void population.
At low temperatures, the removal of small unstable or equilibrium defect clusters reduces the number of
recombination centers, suppresses damage annihilation,  
and speeds the formation of large, stable voids.
This enhances low temperature swelling.
At high temperatures, this same coalescence of small clusters 
greatly reduces the total number of helium bubbles and voids, 
so that the total void sink strength is kept small and  the asymptotic swelling rate 
is diminished compared to the monomer aggregation model (Fig.~\ref{SwellCoal}).
Small clusters are absorbed as rapidly as new ones form, which
prevents the formation of a bimodal distribution of small equilibrium bubbles and large voids.
These differences  suggest that competition between trapping and coalescence of very small (mostly TEM-invisible) clusters significantly shapes the microstructure in real irradiated materials.

When coalescence is included, 
the terminal void density and swelling rate remain sensitive to $\gamma$ up to 500 C.
The predicted void density at this temperature increases from 
$7\times10^{19}$ m$^{-3}$ for $\gamma = 0.75\gamma_0$ 
to  $7\times10^{20}$ m$^{-3}$ for $\gamma = 0.5\gamma_0$.
The swelling rate for the former case is only 0.3\%/dpa at 50 dpa but reaches 0.8\%/dpa for the latter.
This suggests that either the cavity surface energy is substantially smaller than the value for the clean metal or that the vacancy clusters have much smaller mobilities than are modeled here.
The swelling behavior finally becomes insensitive to the surface energy by 700C.
In this limit, coalescence reduces the terminal void density to 4-5$\times10^{18}$ m$^{-3}$.


\section{Conclusions}

This paper introduces a mixed Master Equation/Monte Carlo treatment of rate theory calculations 
in a mean field continuum approximation.
This  enables flexible treatment of the defect density variables, 
using different algorithms to treat the various reactions as efficiently as possible. 
The approximately quasi-stationary distribution of small, unstable or transient clusters is treated 
(as much as possible) by solving continuum rate equations.
This eliminates the need to evaluate rapid individual reactions that mostly cancel one another.
Larger clusters are treated by Monte Carlo methods, which treats clusters of arbitrary size and composition without the need for a fixed grid or artificial discretization of the defect distribution.
A Markov method for smaller clusters accurately simulates rapid fluctuations in size and in the reaction parameters, and a Poisson-distributed random walk efficiently treats the more gradual evolution of the largest clusters.
Finally, a macroparticle approach is introduced to encompass large differences in species densities in the Monte Carlo distribution.

This hybrid scheme readily treats void/bubble evolution to high cumulative fluxes
for temperatures and dose rates that are characteristic of real reactor systems.
Calculations  demonstrate that void coalescence provides an important channel for consolidating vacancy defects into large, stable voids,
controlling the duration of incubation and the 
terminal void density.
It is expected that thermal and radiation-induced micro-chemical  evolution of solute and precipitate distributions will influence the cluster mobility and thereby the macroscopic incubation and steady-swelling behavior.
Some degree of void/bubble trapping seems to be required in order to obtain a bimodal bubble/void size distribution, while some coalescence may be needed to give a 
realistically low terminal void density at higher temperature.
The cavity surface energy determines the barrier for nucleation of stable voids and hence also affects the incubation behavior;  this contribution becomes temperature- and time-dependent if oxygen is explicitly modeled.
All of these effects can be addressed, in principle, by extensions of the method described here.

These calculations also suggest the importance of additional, competing processes that are 
not evaluated at present,
such as interstitial-interstitial aggregation or cluster annihilation from void-dislocation reaction.
The methods described here can be extended to treat coalescence of loops as easily as voids, 
given a suitable binary reaction kernel.
Such reactions should be included for reasons of consistency, 
besides their likely contribution to transient and steady swelling behavior.
They would be especially important if radiation damage were introduced as a variety 
of pre-formed defect clusters.
Based on the preliminary findings for cavity coalescence, 
more general defect cluster reactions are expected to have a significant influence on radiation swelling behavior.

\section{Acknowledgements}

This work performed under the auspices of the U.S. Department of Energy by Lawrence Livermore National Laboratory under Contract DE-AC52-07NA27344.   MPS acknowledges V.V. Bulatov for an early introduction to Markov chain Monte Carlo methods, e.g., Ref.~\cite{CAI:1999}.

\newpage
\bibliographystyle{elsart-num}
\bibliography{CoalMethods4}

\begin{thebibliography}{10}
\expandafter\ifx\csname url\endcsname\relax
  \def\url#1{\texttt{#1}}\fi
\expandafter\ifx\csname urlprefix\endcsname\relax\def\urlprefix{URL }\fi

\bibitem{CAWTHORNE:1967}
C.~Cawthorne, E.~Fulton, Nature 216 (1967) 575.

\bibitem{GarnerWolfer:1984}
F.~A. Garner, W.~G. Wolfer, Factors which determine the swelling behavior of
  austenitic stainless steels, Journal of Nuclear Materials 122-123 (1984)
  201--206.

\bibitem{Okita:2000}
T.~Okita, T.~Kamada, N.~Sekimura, Effects of dose rate on microstructural
  evolution and swelling in austenitic steels under irradiation, Journal of
  Nuclear Materials 283--287 (2000) 220--223.

\bibitem{Okita:2002}
T.~Okita, T.~Sato, N.~Sekimura, F.~A. Garner, L.~R. Greenwood, The primary
  origin of dose rate effects on microstructural evolution of austenitic steels
  during neutron irradiation, Journal of Nuclear Materials 307 (2002) 322--326.

\bibitem{Surh:2004}
M.~P. Surh, J.~B. Sturgeon, W.~G. Wolfer, Master equation and {Fokker-Planck}
  methods for void nucleation and growth in irradiation swelling, Journal of
  Nuclear Materials 325 (2004) 44--51.

\bibitem{Surh:2004b}
M.~P. Surh, J.~B. Sturgeon, W.~G. Wolfer, Vacancy cluster evolution and
  swelling in irradiated 316 stainless steel, Journal of Nuclear Materials 328
  (2004) 107--114.

\bibitem{Surh:2005}
M.~Surh, J.~Sturgeon, W.~Wolfer, Radiation swelling behavior and its dependence
  on temperature, dose rate, and dislocation structure evolution, Journal of
  Nuclear Materials 336 (2005) 217--224.

\bibitem{Surh:ERR}
M.~P. Surh, J.~B. Sturgeon, W.~G. Wolfer, Erratum for: Radiation swelling
  behavior and its dependence on temperature, dose rate, and dislocation
  structure evolution, Journal of Nuclear Materials 328 (2004) 107--114.

\bibitem{Garner:1998}
F.~A. Garner, Irradiation performance of cladding and structural steels in
  liquid metal reactors, in: B.~Frost (Ed.), Materials Science and Technology:
  A Comprehensive Treatment, Vol. 10 A, VCH Verlagsgesellschaft mbH, 1998, pp.
  419--543.

\bibitem{Okita:2001b}
T.~Okita, T.~Sato, N.~Sekimura, F.~A. Garner, The effect of dose rate on
  microstructural evolution in austentic steels irradiated with fast neutrons,
  in: Proceedings of the Fourth Pacific Rim International Conference on
  Advanced Materials and Processing, Vol. PRICM-4, Aoba,Aramaki,Aoba-ku,Sendai,
  2001.

\bibitem{GREENWOOD:1963}
G.~W. Greenwood, M.~V. Speight, An analysis of the diffusion of fission gas
  bubbles and its effect on the behavior of reactor fuels, Journal of Nuclear
  Materials 10 (1963) 140--144.

\bibitem{GRUBER:1967}
E.~Gruber, Calculated size distribution for gas bubble migration and
  coalescence in solids, Journal of Applied Physics 38 (1967) 243--250.

\bibitem{COGHLAN:1983}
L.~K. Mansur, W.~A. Coghlan, Mechanisms of helium interaction with radiation
  effects in metals and alloys: A review, Journal of Nuclear Materials 119
  (1983) 1--25.

\bibitem{MARCUS:1968}
A.~H. Marcus, Stochastic coalescence, Technometrics 10 (1968) 133--143.

\bibitem{GILLESPIE:1972}
D.~Gillespie, The stochastic coalescence model for cloud droplet growth,
  Journal of Atmospheric Science 29 (1972) 1496--1510.

\bibitem{LUSHNIKOV:1978}
A.~A. Lushenko, Some new aspects of coagulation theory, Izvestiya, Akademii
  Nauk SSSR, Fizika Atmosfery i Okeana 14 (1978) 738--743.

\bibitem{VODE}
P.~Brown, G.~Byrne, A.~Hindmarsh, Vode: A variable coefficient ode solver, SIAM
  Journal of Scientific Stat. Comput. 10 (1989) 1038--1051.

\bibitem{SMITH:1998}
M.~Smith, T.~Matsoukas, Constant-number monte carlo simulation of population
  balances, Chemical Engineering Science 53 (1998) 1777--1786.

\bibitem{BABOVSKY:1999}
H.~Babovsky, On a monte carlo scheme for smoluchowski's coagulation equation,
  Monte Carlo Methods and Applications 5 (1999) 1--18.

\bibitem{GILLESPIE:2000}
D.~Gillespie, The chemical langevin equation, Journal of Chemical Physics 114
  (2000) 297--306.

\bibitem{GILLESPIE:2001}
D.~Gillespie, Approximate accelerated stochastic simulation of chemically
  reacting systems, Journal of Chemical Physics 115 (2001) 1716--1733.

\bibitem{EIBECK:2001}
A.~Eibeck, W.~Wagner, Stochastic particle approximations for smoluchowski's
  coagulation equation, The Annals of Applied Probability 11 (2001) 1137--1165.

\bibitem{HASELTINE:2002}
E.~L. Haseltine, J.~B. Rawlings, Approximate solution of coupled fast and slow
  reactions for stochastic chemical kinetics, Journal of Chemical Physics 117
  (2002) 6959--6969.

\bibitem{FRIESEN:2003}
W.~I. Friesen, T.~Dabros, Constant-number monte carlo simulation of aggregating
  and fragmenting particles, Journal of Chemical Physics 119~(5) (2003)
  2825--2839.

\bibitem{LAURENZI:2003}
I.~I. Laurenzi, S.~L. Diamond, Kinetics of aggregation-fragmentation processes
  with multiple components, Physical Review E 67 (2003) 051103--1--15.

\bibitem{MUKHERJEE:2003}
D.~Mukherjee, C.~Sonwane, M.~Zacariah, Kinetic monte carlo simulation of the
  effect of coalescence energy release on the size and shape evolution of
  nanoparticles grown as an aerosol, Journal of Chemical Physics 119 (2003)
  3391--3403.

\bibitem{ALEXOPOULOS:2004}
A.~H. Alexopoulos, A.~I. Roussos, C.~Kiparissides, Dynamic evolution of the
  particle size distribution in particulate processes undergoing combined
  particle growth and aggregation, Chemical Engineering Science 59 (2004)
  5751--5769.

\bibitem{FILBET:2004}
F.~Filbet, P.~Laurencot, Numerical simulation of the smoluchowski coagulation
  equation, SIAM Journal of Scientific Computing 25~(6) (2004) 2004--2028.

\bibitem{SALIS:2005}
H.~Salis, Y.~Kaznessis, Accurate hybrid stochastic simulation of a system of
  coupled chemical or biochemical reactions, Journal of Chemical Physics 122
  (2005) 054103.1--13.

\bibitem{KRAFT:2005}
M.~Kraft, Modelling of particulate processes, KONA, Powder and Particle 23
  (2005) 18--35.

\bibitem{WOLFER:1985}
W.~G. Wolfer, B.~B. Glasgow, Dislocation evolution in metals during
  irradiation, Acta Metallurgica 33 (1985) 1997--2004.

\bibitem{SNIEGOWSKI:1983}
J.~J. Sniegowski, W.~G. Wolfer, On the physical basis for the swelling
  resistance of ferritic steels, in: J.~W. Davis, D.~J. Michel (Eds.),
  Proceedings of Topical Conference on Ferritic Alloys for Use in Nuclear
  Energy Technologies, Snowbird, Utah, 1983, pp. 579--586.

\bibitem{SurhWolfer:TBP}
M.~P. Surh, W.~G. Wolfer, Accurate mean field void bias factors for radiation
  swelling calculations, Journal of Computer Aided Design 14 (2007) 419--424.

\bibitem{GARNER:2003}
L.~Greenwood, F.~Garner, B.~Oliver, M.~Grossbeck, W.~G. Wolfer, Surprisingly
  large generation and retention of helium and hydrogen in pure nickel
  irradiated at high temperatures and high neutron exposures, in: M.~L.
  Grossbeck, T.~R. Allen, R.~G. Lott, A.~S. Kumar (Eds.), The Effects of
  Radiation on Materials: 21st International Symposium, ASTM STP 1447, American
  Society for Testing and Materials International, West Conshohocken, PA, 2003.

\bibitem{SCHALDACH:2003}
C.~Schaldach, W.~Wolfer, Kinetics of helium bubble formation in nuclear and
  structural materials, in: M.~L. Grossbeck, T.~R. Allen, R.~G. Lott, A.~S.
  Kumar (Eds.), The Effects of Radiation on Materials:21st International
  Symposium, ASTM STP 1447, American Society for Testing and Materials
  International, West Conshohocken, PA, 2003.

\bibitem{MOONPAO:1967}
F.~C. Moon, Y.~H. Pao, Journal of Applied Physics 38 (1967) 595.

\bibitem{WOLFERASHKIN:1973}
W.~G. Wolfer, M.~Ashkin, Scripta Metallurgica 7 (1973) 1175.

\bibitem{COTTRELL:2002}
G.~Cottrell, Void migration in fusion materials, Journal of Nuclear Materials
  302 (2002) 220--223.

\bibitem{Wolfer:1988}
W.~G. Wolfer, The pressure for dislocation loop punching by a single bubble,
  Philosophical Magazine A 58 (1988) 285--297.

\bibitem{ADAMS:1989}
J.~Adams, W.~Wolfer, Formation energies of helium-void complexes in nickel,
  Journal of Nuclear Materials 166 (1989) 235--242.

\bibitem{English:1987}
C.~A. English, B.~L. Eyre, J.~W. Muncie, Low-dose neutron irradiation damage in
  copper ii. damage-structure evolution at elevated temperatures, Philosophical
  Magazine A 56 (1987) 453--484.

\bibitem{WILSON:1983}
W.~D. Wilson, Theory of small clusters of helium in metals, Radiation Effects
  78 (1983) 11--24.

\bibitem{NELSON:1966}
R.~S. Nelson, On the binding of inert gas bubbles to precipitates, Journal of
  Nuclear Materials 19 (1966) 149--154.

\bibitem{MIKHLIN:1979}
E.~Mikhlin, Suppression of diffusion mobility of small gas bubbles in solids,
  Physica Status Solidi (a) 56 (1979) 763--768.

\bibitem{ALEXANDER:1992}
D.~E. Alexander, R.~C. Birtcher, The effect of ion irradiation on inert gas
  bubble mobility, Journal of Nuclear Materials 191--194 (1992) 1289--1294.

\bibitem{GOLUBOV:2001}
S.~I. Golubov, A.~M. Ovcharenko, A.~V. Barashev, B.~N. Singh, Grouping method
  for the approximate solution of a kinetic equation describing the evolution
  of point defect clusters, Philosophical Magazine A 81~(3) (2001) 643--658.

\bibitem{SOLOVYEV:1999}
A.~Solovyev, V.~Terekhin, V.~Tikhonchuk, L.~Altgilgers, Electron kinetic
  effects in atmosphere breakdown by an intense electromagnetic pulse, Physical
  Review E 60 (1999) 7360--7368.

\bibitem{MACRO}
C.~Birdsall, A.~Langdon, H.~Okuda, Finite-size particle physics applied to
  plasma simulation, in: B.~Alder, S.~Fernbach, M.~Rotenberg (Eds.), Methods in
  Computational Physics, Vol.~9, Academic Press, New York, 1970, pp. 241--258.

\bibitem{MACKEOWN}
P.~MacKeown, Stochastic Simulation in Physics, Springer-Verlag, Singapore,
  1997.

\bibitem{NUMREC3}
W.~H. Press, B.~P. Flannery, S.~A. Teukolsky, W.~T. Vetterling, Numerical
  Recipes: The Art of Scientific Computing, Cambridge University Press, 1986.

\bibitem{SABLEFELD:1996}
K.~Sablefeld, S.~Rogasinsky, A.~Kolodko, A.~Levykin, Stochastic algorithms for
  solving the smolouchovsky coagulation equation and application to aerosol
  growth simulation, Monte Carlo Methods and Applications 2 (1996) 41--87.

\bibitem{NUMREC5}
W.~H. Press, B.~P. Flannery, S.~A. Teukolsky, W.~T. Vetterling, Numerical
  Recipes: The art of Scientific Computing, Cambridge University Press, 1986.

\bibitem{MANSUR:1986}
E.~H. Lee, L.~K. Mansur, A mechanism of swelling suppression in
  phosphorus-modified fe-ni-cr alloys, Journal of Nuclear Materials 141--143
  (1986) 695--702.

\bibitem{MANSUR:1990}
E.~H. Lee, L.~K. Mansur, Mechanisms of swelling suppression in cold-worked
  phosphorus-modified fe-ni-cr alloys, Journal of Nuclear Materials 61 (1990)
  733--749.

\bibitem{WEHNER:1985}
M.~F. Wehner, W.~G. Wolfer, Vacancy cluster evolution in metals under
  irradiation, Philosophical Magazine A 52 (1985) 189--205.

\bibitem{CAI:1999}
W.~Cai, V.~Bulatov, S.~Yip, Kinetic monte carlo method for dislocation glide in
  silicon, Journal of Computer-Aided Materials Design 6 (1999) 175--183.

\end{thebibliography}
\newpage

%

%

%

%

\newpage
\begin{table}[htdp]
\caption{Model material parameters for type-316 stainless steel. }
\begin{center}
\begin{tabular}{|p{3.5cm}|c|c|}
\hline
\multirow{6}{*}{\vbox{Bulk parameters:}} 
&Lattice constant $a_0$                      & $3.639\times10^{-10}$ m       \\
&Burgers vector $b$                                & $a_0/\sqrt{2}$                        \\
&Atomic volume $\Omega$                    & ${a_0}^3/4$                           \\
&Shear modulus $\mu$            & $8.295\times10^{10}$ Pa                  \\
&Poisson's ratio $\nu$              & 0.264                                                      \\
&Cascade efficiency $\xi_{Frenkel}$                      & 0.1                      \\
\hline
\multirow{6}{*}{Vacancy parameters:}
&Relaxation volume               & -0.2  $\Omega$                             \\
&Migration energy   $E_{m}$    & $2.08\times10^{-19}$ J          \\
&Formation energy $E_f$      & $2.88\times10^{-19}$ J               \\
&Formation entropy $S_f$   & 1.5 $k_B$                                        \\
&Pre-exponential factor                     & $1.29\times10^{-6}$   m$^2$/s      \\
&Shear polarizability                        & $ -2.4\times10^{-18}$               \\
\hline
\multirow{4}{*}{\vbox{Self-interstitial \\ parameters:}}
&Relaxation volume              & 1.50    $\Omega$                          \\
&Migration energy $E_m$   & $0.320\times10^{-19}$  J             \\ 
&Pre-exponential factor                        & $1.29\times10^{-6}$  m$^2$/s     \\
&Shear polarizability                      & $-2.535\times10^{-17} $           \\
\hline
\multirow{4}{*}{\vbox{Interstitial helium \\ parameters:}} 
&Relaxation volume              & 0.60    $\Omega$                      \\
&Migration energy $E_m$   & $0.320\times10^{-19}$ J          \\ 
&Pre-exponential factor                        & $1.29\times10^{-6}$  m$^2$/s             \\
&Shear polarizability                      & $-2.535\times10^{-17} $           \\
\hline
\multirow{4}{*}{\vbox{Substitutional helium \\ parameters:}}
&Relaxation volume               & -0.2  $\Omega$                \\
&Migration energy   $E_{m}$    & $2.08\times10^{-19}$ J      \\
&Pre-exponential factor                     & $1.29\times10^{-6}$   m$^2$/s              \\
&Shear polarizability                        & $ -2.4\times10^{-18}$               \\
\hline
\multirow{5}{*}{Void parameters:} 
&Relaxation volume               &  0                                                                      \\
&Surface energy  $\gamma_0(T=0)$              & 2.408  J/m$^2$                          \\
&Temperature derivative $d\gamma_0/dT$        & $0.440\times10^{-3}$  J/m$^2$/K   \\
& Adsorption factor $\Lambda$ & 0.45-0.80\\
&Migration energy   $E_{m}$    & $2.08\times10^{-19}$ J                         \\
&Pre-exponential factor                     & $1.29\times10^{-6}$   m$^2$/s  /$n_{v}^{4/3}$      \\
\hline
\multirow{3}{*}{\vbox{Environmental \\parameters:}} 
&Temperature   $T$            &  300-700 C                                                                     \\
&Flux   $\phi$            &  $10^{-6}$ dpa/s                                                                      \\
&Damage efficiency   $\xi$            &  0.1                                                                      \\
\hline

\end{tabular}
\end{center}

\label{TableOne}
\end{table}


\end{document}